\documentclass[fleqn,usenatbib]{mnras}

\usepackage{newtxtext,newtxmath}

\usepackage[T1]{fontenc}

\DeclareRobustCommand{\VAN}[3]{#2}
\let\VANthebibliography\thebibliography
\def\thebibliography{\DeclareRobustCommand{\VAN}[3]{##3}\VANthebibliography}

\usepackage{graphicx}	
\usepackage{amsmath}	
\usepackage{tabularx,ragged2e}
\usepackage{balance}

\newcolumntype{Y}{>{\RaggedRight\arraybackslash}X}

\title[Action Bar Classification ]{Bar Properties and Star-by-Star Bar Membership via Action Conservation}

\author[E.J. Iles et al.]{
Elizabeth J. Iles,$^{1,2}$\thanks{E-mail: elizabeth.iles@sydney.edu.au)}
Finn A. Pal,$^{2}$
Joss Bland-Hawthorn,$^{1}$
and Ken Freeman$^{3}$
\\
$^{1}$Sydney Institute for Astronomy (SIfA), Sydney University, Camperdown 2050, Australia\\
$^{2}$School of Physics, University of New South Wales, Kensington 2052, Australia\\
$^{3}$Research School of Astronomy \& Astrophysics, Mount Stromlo Observatory, Cotter Road, Weston Creek, ACT 2611, Australia
}

\date{Accepted XXX. Received YYY; in original form ZZZ}

\pubyear{2026}

\begin{document}
\label{firstpage}
\pagerange{\pageref{firstpage}--\pageref{lastpage}}
\maketitle

% Abstract of the paper
\begin{abstract}
A bar-like central feature is commonly observed in both nearby and distant spiral-type galaxies, including the Milky Way. While many methods exist to categorise this morphology, no one method has emerged as the field-wide standard. To develop a rigorous and consistent method for identifying these bars, we investigate a classification scheme based on dynamical actions. In the Gaia era, actions can be estimated for individual stars in both observations and simulations, making this a natural and unifying diagnostic, assuming the intrinsic errors and selection functions are understood. Our approach is straightforward: stars that participate in the bar are subject to a strongly non-axisymmetric potential and, therefore, do not completely conserve their actions. We use this property to define a star-by-star criterion, formulated as an inequality and evaluated within measurement uncertainties, to identify bar members based on the degree to which their total action fails to be conserved. From tests on simulated galaxies, we find that the bar region is indeed characterised by a lower fraction of stars with conserved actions and that stars on bar orbits are represented by larger percentage changes in their actions. We are able to classify the spatial extent of barred region via the standard parameters of bar length and orientation, while also individually separating bar-located from bar-member stars on bar orbits. As proof of concept, our automated method based on dynamical actions robustly identifies bar parameters that closely match the eye's performance (average bar length variation $\sim9$\%) in barred snapshots of the test galaxy. 

\end{abstract}

%%%%%%%%%%%%%%%%%%%%%%%%%%%%%%%%%%%%%

%%%%%%%%%%%%%%%%%%%%%%%%%%%%%%%%%%%%%%%%%%%%%%%%%%

%%%%%%%%%%%%%%%%% BODY OF PAPER %%%%%%%%%%%%%%%%%%

\section{Introduction}
\subsection{Background}
The ubiquitous bar-like structure present in the centre of many disk galaxies, while a key morphological feature for galaxy classification, is similarly significant for driving galaxy evolution. Bars have been demonstrated to affect host-galaxy processes on all scales: from star formation on molecular-cloud scales \citep[$10-100$\,pc; e.g.][]{Kuno2000, Dobbs2006, Dobbs2014, Nilipour2024, Maeda2025} to galaxy-wide dynamics \citep[$\sim1000$\,pc; e.g.][]{Regan1999, ONeill2003, Emsellem2006, iles2024, Geron2024, Rutherford2025}; including disk–halo and disk–bulge co-evolution \citep[e.g.][]{Athanassoula2002, Valenzuela2003, Kormendy2004, Jogee2005, Athanassoula2013, Kruk2018}, angular momentum distribution \citep[e.g.][]{Weinberg1985, Athanassoula2005, Romeo2023}, Active Galactic Nuclei (AGN) \citep[e.g.][]{Shlosman1989, Alonso2014, Garland2023}, Inter-Stellar Medium (ISM) structure \citep{Friedli1993, Lindblad1994, Pettitt2020, Katsioli2026}, gas depletion and quenching \citep[][]{Masters2012, Spinoso2017, Fraser-McKelvie2020, Geron2021}, for example. Accounting for up to $\sim75$\% of nearby disk galaxies \citep[e.g.][]{schinnerer2002, Aguerri2009, Nair2010, Saha2018}, an increasing number of galaxies at high redshift \citep[e.g.][]{Guo2023, Costantin2023, Liang2024, LeConte2024, LeConte2025}, and even the Milky Way \citep[e.g.][]{Blitz1991, Nakada1991, Paczynski1994, Zhao1994, Bland-Hawthorn2016}, these features are well-studied and yet, despite their ubiquity and impact, we still lack a distinct, universally-applicable method used to define the spatial extent of the “bar” in these galaxies \citep[see, for recent example: ][]{Dehnen2023,Lucey2023,Petersen2024,iles2025}, let alone to identify whether an individual star is a member of the bar 
or just passing through. 
For example, Fourier analysis \citep[e.g.][]{Elmegreen1985, Ohta1990, Aguerri1998, Garcia-Gomez2017, Pettitt2018, Frosst2025} has long been used to identify bars, particularly in simulations, while observational studies have traditionally relied on isophotal ellipse fitting \citep[e.g.][]{Abraham1999, Laine2002, Erwin2005, Gadotti2006, Marinova2007, Menendez-Delmestre2007, Consolandi2016, Jiang2018} or photometric decomposition \citep[e.g.][]{Reese2007, Gadotti2008, Durbala2008, Durbala2009, Weinzirl2009, Kruk2018, Fraser-McKelvie2025}. Visual classification \citep[e.g.][]{Nair2010, Hoyle2011, Masters2011, Masters2012, Melvin2014, Kruk2018, Masters2021, Geron2021} remains common, as bars are inherently visual features, but this is increasingly impractical for large surveys and is inherently subject to significant person-to-person variability \citep[e.g.][]{Lahav1995, Heidl2013, Qian2022, iles2025}. Crowdsourcing \citep[e.g.][]{Lahav1995, Fortson2018, Foster2025} can reduce this variation but can be similarly inefficient and subjective if not applied carefully. More recently, automated bar-detection methods using non‑parametric deep learning, via classification, regression, and segmentation \citep[e.g.][]{Abraham2018, Cavanagh2020, Fluke2020, Cavanagh2022, Huertas-Company2023, Cavanagh2024}, have become increasingly prevalent. However, since there are so many feasible methods to identify the bar in galaxies, it is necessary to select a `best’ method for each study.  This presents a fundamental problem: do any of these methods consistently recover the same structure across all barred galaxies?

The definition of a bar, or barred region, is fundamental to many studies of host-galaxy properties, such as gas flow or star formation along the bar \citep[e.g.][]{Downes1996, Sheth2002, Momose2010, Casasola2011, Watanabe2019, Querejeta2021, iles2022}, as well as correlations between bar length and large-scale galactic properties, such as longer bars in more massive galaxies \citep[e.g.][]{Kormendy1979, Erwin2005, Diaz-Garcia2016}, or shorter bars in late-type galaxies \citep[e.g][]{Elmegreen1985, Combes1993, Erwin2005, Aguerri2009, Erwin2019} and whether this may be redshift dependent \citep[e.g.][]{Sheth2008, Kim2021}. If the definition of the bar differs between methods, the physical trends we observe may be significantly effected by this systematic bias. So, despite our ability to define the bar in many ways, defining the bar in a way that is distinct, repeatable, and applicable across barred galaxies must still be a priority for the field. It is only in this way that we can most effectively approach a unified understanding of bar structure, galaxy formation, and the role of bars in driving galaxy evolution.

\subsection{Dynamical Actions}
We propose a different approach to classifying the bar in galaxies, particularly aimed at identifying bars in simulations with high resolution capturing complex or non-standard systems, such as double bars, nuclear disks, or bars distorted by interactions, with the added capacity to track time evolution on galactic scales. Further, this process is designed be able to determine, not only the spatial extent of the bar but also, the membership of individual stars to the bar, distinguishing stars on bar-supporting orbits, or orbits affected by the bar, from those merely passing through. This capability will enable the assessment of important bar-driven processes, including star formation, radial migration and chemical mixing, on a more precise, star-by-star basis, at least in simulations. 

The basis of this method is to exploit the intrinsic properties of the angle–action variables $(\theta_i, J_i)$ in different regions of the galactic potential. Action-space, described by these angle-action variables, has long been used to probe many properties of the Milky Way through Galactic dynamics \citep[e.g.][]{Lynden-Bell1972, Kalnajs1977, Arnold1978, Martinet1981, Rauch1996, Weinberg2001, Sellwood2012}. This is primarily due to the convenience of the action integrals, as a set of adiabatic invariant quantities for any quasi-periodic orbit in a steady, or slowly varying, potential \citep[see][for a more specific detail]{Binney2008}. Actions have been instrumental in modelling phase-mixed populations \citep[e.g.][]{Tremaine1999, Antoja2018, Ting2019, Monari2019, Drimmel2023}, identifying stellar streams \citep[e.g.][]{Helmi1999, Eyre2011, Sanders2013}, and better constraining the dark matter distribution within the Galaxy \citep[e.g.][]{Machado2016, Malhan2022, Carlberg2024}. Most importantly, the more abstract action-angle co-ordinates can be easily related to ordinary phase-space variables $(x, v)$ \citep[e.g.][]{Binney2008, Binney2020}, either directly from the stellar orbits or via the instantaneous position and velocity measurements with an approximation of a local St{\"a}ckel potential \citep[e.g.][]{Binney2012, Sanders2015}, and algorithms that compute actions are readily available, relatively widespread and mostly computationally inexpensive, practical even for million‑particle $N-$body simulations or extremely large observational datasets  \citep[e.g.][]{Debattista2020, Mikkola2020, Zozulia2024}. 

The most useful invariant property of the action, however, relies primarily on the assumption that the potential is axisymmetric and invariant. 
Actions cease to be conserved in non-axisymmetric environments, such as bars and spiral arms \citep[e.g.][]{Binney2008, Solway2012, Vera-Ciro2016}; resonances, such as co-rotation \citep[CR; e.g.][]{Sellwood2002, Roskar2012, Daniel2015} and the inner- and outer-Lindblad resonances \citep[ILR/OLR; e.g.][]{Barbanis1967, Lynden-Bell1972, Carlberg1985}, can significantly alter individual actions; vertical action may be conserved during radial migration in general, but not necessarily on a star-by-star basis \citep[e.g.][]{Solway2012, Vera-Ciro2016, Mikkola2020}; spiral perturbations can change azimuthal action while leaving radial action nearly unchanged \citep[e.g.][]{Sellwood2002, Roskar2012}; and rapid potential evolution, such as during bar formation and buckling \citep[e.g.][]{Toomre1981, Sellwood1981, Thielheim1982, Raha1991}, can violate adiabatic invariance entirely. Assumptions and approximations have sought to mitigate the effects of this variance to probe galactic history \citep[][]{McMillan2011, Binney2012, Sanders2015, Antoja2018, Ting2019, Malhan2022, Carlberg2024}, while recent studies have started to explore how resonances and the action distribution of stars are related to bars, spirals, and boxy/peanut bulges \citep[e.g.][]{Debattista2020, Kawata2021, Trick2021, Trick2022, Drimmel2023, Zozulia2024, Arunima2025}. In this work, we are only concerned with exploiting the existence of these changes to the action, rather than reducing or categorising their effects. The premise here is: if actions are systematically not conserved for bar‑located stars but remain broadly conserved elsewhere in the disc, then the non‑conservation itself becomes a diagnostic. This will allow us to directly infer bar membership and identify bar properties through a simple assessment of the conservation of action for stars throughout the galactic potential.

In this work, we demonstrate this method in the context of numerical $N-$body simulations, making the best use of simulated data to track time-evolution. In a companion paper \citep[see also][]{iles2026} we demonstrate the application of this method for the Milky Way bar, using observations of stars in large surveys. A description of the method and test simulation is included in Section \ref{s:methods}. The definition of a standard bar region (e.g. radial bar extent and pitch-angle) is included in Section \ref{s:barproperties}, while Section \ref{s:starbystar} outlines the `star-by-star' use-case, identifying potential bar membership for each stellar particle in the simulation. Finally, our conclusions and recommendations for future applications of this method are included in Section \ref{s:conclusion}. 

%%%%%%%%%%%%%%%%%%%%%%%%%%%%%%%%%%%%%%%%%%%%%%%%%%%%%%%%%%%%%%%%%%%%%%
\section{Data \& Methods}
\label{s:methods}
%%%%%%
\subsection{The Test Simulations }
\label{ss:methodsims}
We demonstrate the practicality of the proposed method using $N-$body SPH (smoothed particle hydrodynamics) in the form of an isolated disk galaxy simulation (IsoB) from \citet{iles2022}. To test bar-finding on a star-by-star basis, simulations with high temporal and spatial resolution were necessary to be able to trace bar evolution and stellar orbits. The IsoB simulation has a stellar particle mass resolution of $1044$\,M$_\odot$ for simulation-formed stars, a stellar softening length of 50\,pc and a snapshot for every 10\,Myr of galaxy evolution, which is approximately equivalent to the star formation timescale ($t_{\rm SF} = 10$\,Myr). This disk was evolved with \textsc{gasoline2} \citep{wadsley2017}, comprising a dark matter component, a small stellar bulge, an initial stellar disk and a gas disk component of particles respectively, following the standard sub-grid prescriptions \citep{Katz1996,Stinson2006}; with super-bubble feedback \citep{Keller2014}, UV and photoelectric heating, as well as metal cooling in the form of a tabulated cooling function \citep{Shen2010}, recovering the two-phase ISM \citep[][]{Wolfire2003} from an initially isothermal ($10^4$\,K) gas profile. Star formation was prescribed by a standard temperature (300\,K) and density (100\,atoms/cc) threshold, as well as a convergent flow requirement with an efficiency of $\sim10$\% and a \citet{Chabrier2003} initial mass function (IMF). The initial condition was originally tailored to observations of nearby, resolved galaxy \textsc{ngc4303} via the rotation curve, gas and stellar mass fractions \citep[see ][ for more details]{iles2022}. The simulation was evolved for a relatively short period of $1$\,Gyr, however, this is sufficient to cover a period pre-bar formation, the initial bar formation, and a short period post-bar formation to capture action properties in each of these regimes. Later snapshots of the IsoB disk can be considered a classic example of a barred galaxy, however, automated bar identification methods do not consistently produce realistic bar parameters for this disk in every snapshot. 
Thus, we consider this simulation a good elementary demonstration of 
the proposed method, as there are sufficient snapshots with automated bar classification for comparison but automation 
can 
struggle with this galaxy. Additionally, we have recently completed visual classification for each snapshot of the IsoB simulation for a study into visual bar classification \citep[][]{iles2025} which, despite some limitations, can also be used as a comparison for the bar parameters of this disk as measured `by-eye'. 

%%%%%%
\subsection{Calculating the Action of Star Particles}
\label{ss:methodactions}

\begin{figure}
    \centering
	\includegraphics[width=.97\columnwidth]{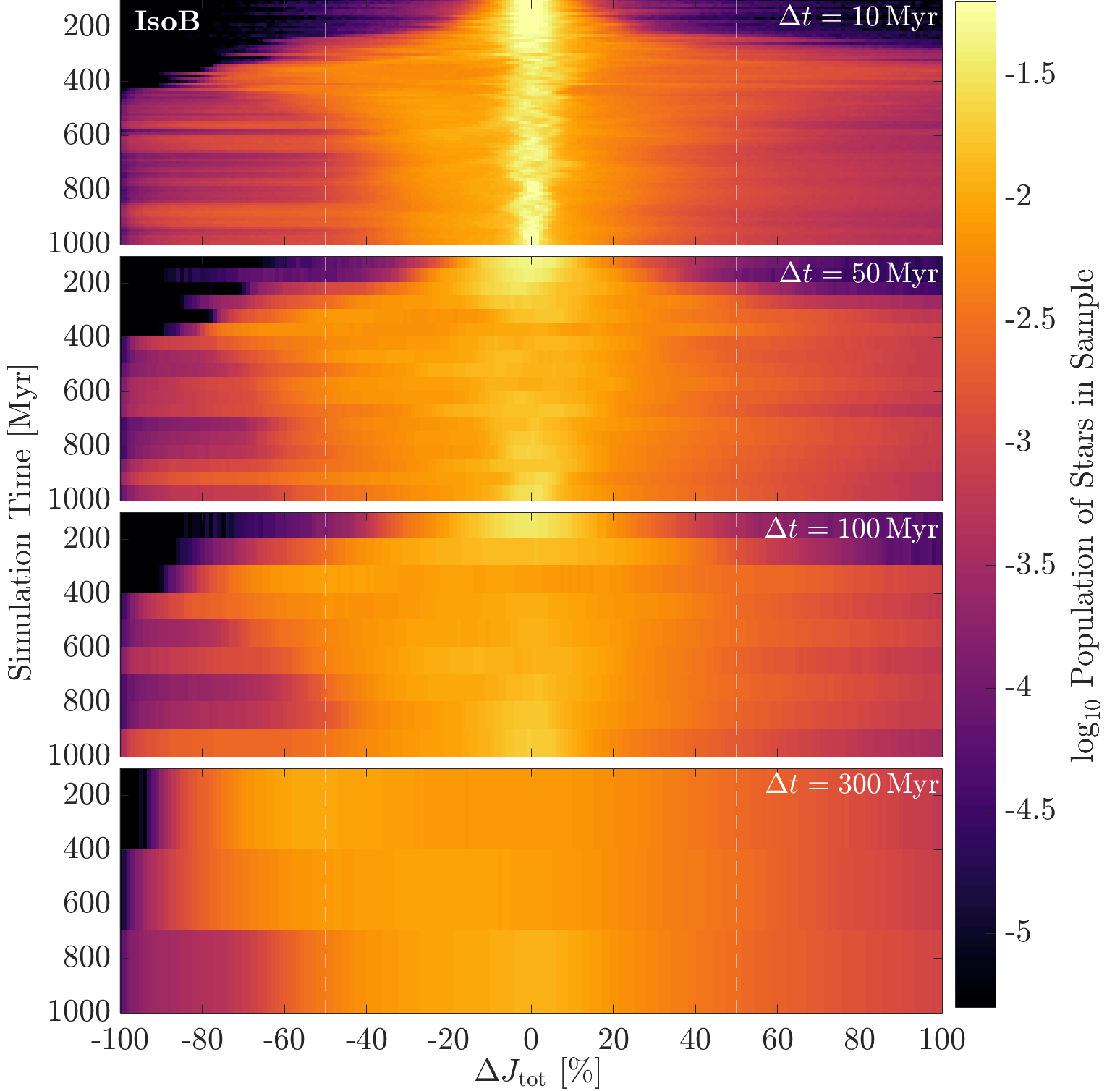}
    \caption{The percentage change in total action $J_{\rm tot}$ for time periods $\Delta t = 10, 50, 100, 300$\,Myr increasing downward in each panel. The colour is a measure of the number of star particles existing at both the start and end of the period measured (dark = less, light = more) and the scale is logarithmic. This is the isolated simulation (IsoB). Fine details obvious in the upper panel increasingly blur in lower panels.}
    \label{f:dJtot_dt_hist}
\end{figure}

Initially, we calculated actions for the simulated star particles using the \textsc{galpy} python package for galactic dynamics with an inbuilt interpolation routine to determine an axisymmetrised potential extracted from the $N-$body particles in the simulation \citep{Bovy2015}. Using \textsc{galpy}, the actions are computed under the St{\"a}ckel approximation of \citet{Binney2012}, producing values for each star particle in component form: the radial action ($J_{R}$) broadly describing the eccentricity of the particle’s orbit; the vertical action ($J_{z}$) describing the deviation away from the disk-plane; and the azimuthal action ($J_{\phi}$), also known as the z-component of the angular momentum ($L_z$) and related to the guiding radius of the particle’s orbit. We then combine these action-angle dependent values into an action-space ($\mathbf{J}$) distance metric, 
\begin{equation}
    J_{\rm tot} = \vert \mathbf{J}\vert = \sqrt{J^2_R + J^2_\phi + J^2_z} .
\end{equation}
Any function of the actions is conserved if the system remains exactly integrable and unperturbed.
While, this particular quantity is not a useful dynamical diagnostic for, say, the construction of a distribution function, because the quadrature sum removes the direction (sign) dependence \citep[cf.][]{will14,post15}. But, as we show, our experiments demonstrate this metric is effective at separating orbits that contribute to the bar, and those that do not. We are interested only in a binary outcome - to determine whether the metric is conserved or not conserved for star particles, within some tolerance calibrated by the simulations.

\begin{figure*}
    \centering
    \includegraphics[width=0.9\textwidth]{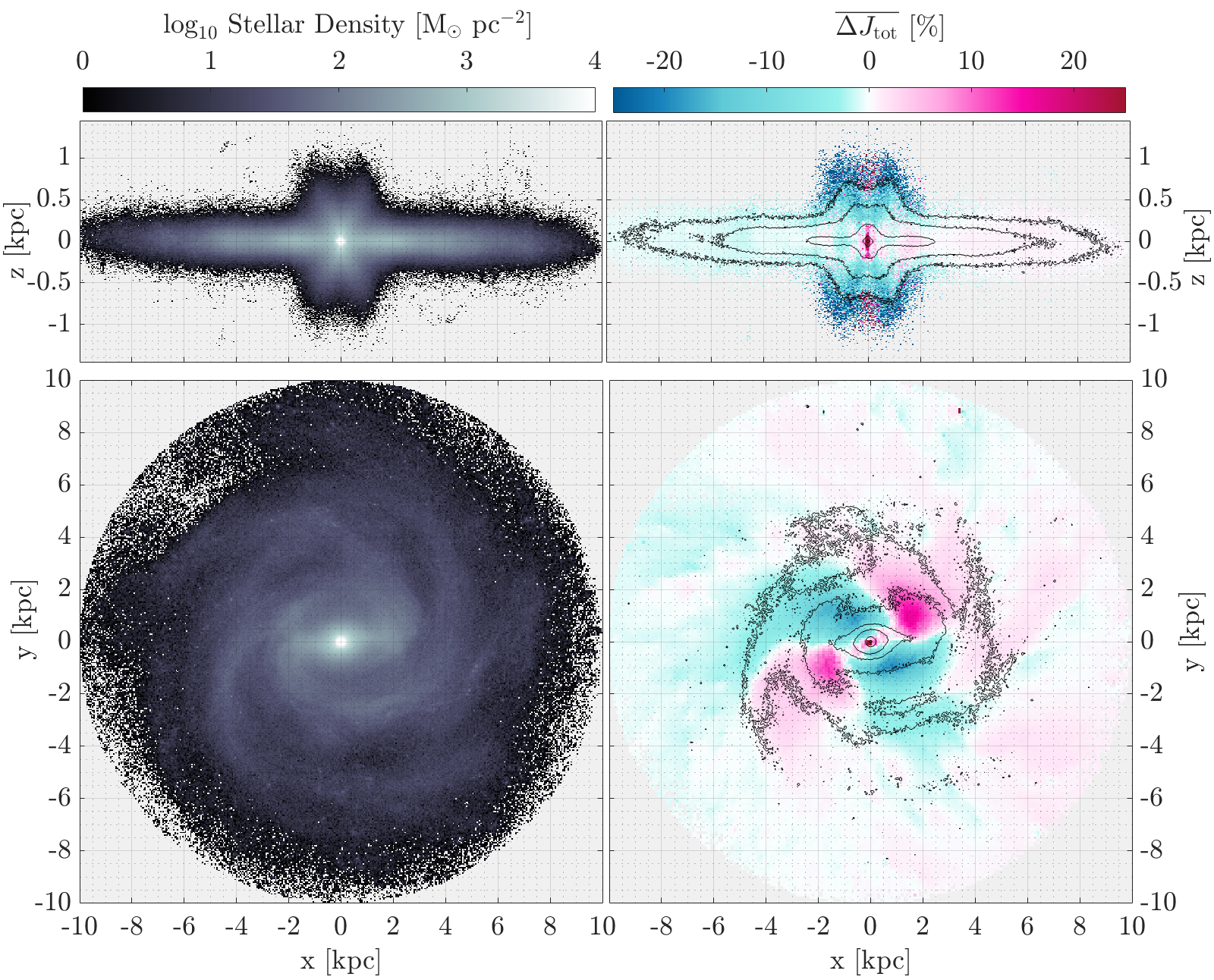}
    \caption{Stellar mass density (left) and average change in action ($\Delta J_{\rm tot}$\,\%) for the $1$\,Gyr snapshot in edge-on ($xz$\,-\,upper) and face-on ($xy$\,-\,lower) orientation. The projected values  are measured along the line-of-sight with a pixel grid of size: $x=y=50$\,pc and $z=10$\,pc and within a total extent of $R<10$\,kpc, $|z|<3$\,kpc. Black contours in the right panel trace stellar density peaks from the left panel.}
    \label{f:staractionprj}
\end{figure*}

Next, if we are to consider a measure of `conservation', this must inherently be a measurement of \emph{change} over some time interval. To determine the most appropriate period ($\Delta t$), we measured the change in total action ($J_{\rm tot}$) over a range of time periods ($\Delta t = 10$, $50$, $100$, $300$\,Myr) which are fractions ($0.01,\ 0.05, \ 0.1, \ 0.3$) of our total simulation time respectively. The results of this test are presented in Figure\,\ref{f:dJtot_dt_hist}. 
This figure is a histogram of the change in total action ($\Delta J_{\rm tot}$) over each $\Delta t$ value (panels, increasing downward) presented as a heat-map at each resolved snapshot of the simulation (every $10$\,Myr), with time also increasing downward along the $y-$axis and starting from $t=100$\,Myr for divisibility, as well as to avoid any settling effects in the start-up of the simulation. On the $x$-axis, the percentage change in total action ($(J_{\rm tot}(t_{\rm final})-J_{\rm tot}(t_{\rm initial}))/J_{\rm tot}(t_{\rm initial})$\%) is used as a value independent metric for categorising the extent of action change. This means that a value of zero indicates no change in the action over the period specified, while a value of 100 indicates a 100\% change in the value of the action. Dashed white lines are used to visually indicate where the total action changes by half the original amount. 
We will use this percentage change in total action ($\Delta J_{\rm tot}$\%) parameter throughout the following analysis. As the level of detail in the colour distribution of each panel becomes increasingly blurred (similar colour spreads across the image) over larger time periods (increasing downward), we select the smallest period ($\Delta t = 10$\,) as the standard for all future analysis. This is because we intend to exploit the conservation of action to constrain disk properties with the highest possible precision. However, by eye it would appear that the $\Delta t \le 100$\,Myr window still retains at least the large-scale features in this representation. Using a simple model of a flat rotation curve at the maximum rotational velocity ($V_{\rm circ}\sim200$\,km s$^{-1}$) and the mid-disk radius ($R\sim 5$\,kpc), we can find an approximate rotational period for the disk ($T_{\rm disk}\sim150$\,Myr), which is likely the limit ($\Delta t \le T_{\rm disk}$) at which structural variation using this method could be acquired. 

Finally, we note that defining the axisymmetric form of the galactic potential is fundamental to the calculation of the actions. 
Thus we also used the \textsc{agama} package \citep{Vasiliev2019} as an alternative method to determine the actions. For the \textsc{agama} potential, we initially modelled the stars and cold gas ($\log T < 4.5$) with an azimuthal harmonic expansion and the dark matter and hot gas with a spherical harmonic expansion. Then, to recover the actions in the non-axisymmetric regime, \textsc{agama} also makes use of the St{\"a}ckel approximation. However, we found that there were a selection of snapshots for which the calculation of actions would fail in these potentials, returning no values. As such, we trialled a different method of construction, where both baryons (stars + gas) and dark matter were instead modelled with spherical harmonic expansions. This approach allowed us to determine actions in a wider range of snapshots, with only a slight drop in accuracy (<1\%) as compared to the initial method with \textsc{agama}. Although both \textsc{galpy} and \textsc{agama} rely on the St{\"a}ckel approximation to navigate the non-axisymmetric conditions, there are slight differences in the calculation method and, we find, in the result (see Figure\,\ref{f:Rbar_actions} in Section\,\ref{s:barproperties}). Finally, for completeness, we tested the simplest option. Instead of integrating or modelling the existing potential from the $N$-body particles, we used standard forms of a Milky Way-like potential and once again recovered the actions using the \textsc{agama} package. In this instance, we used a \citet{Dehnen1993} type bulge, a \citet{MiyamotoNagai1975} potential for the stellar disk, and a NFW \citep{NFW1997} potential for the gas and dark matter. Each of these potentials were scaled using the initial scale radii and masses for each component from \citet{iles2022}, to avoid extra calculations at each snapshot, as this is already an approximation. We do not expect this to be the most accurate representation of the galactic potential, 
however, it is very much the simplest. In comparing the bar parameters we determine using the actions under each potential, we indeed find some differences but the magnitude of these differences can be considered relatively insignificant. We include a discussion of this comparison between potentials alongside our results in Section\,\ref{s:barproperties}.

%%%%%%%%%%%%%%%%%%%%%%%%%%%%%%%%%%%%%%%%%%%%%%%%%%
\section{The ``Bar Region'' in Projection}
\label{s:barproperties}

\begin{figure*} 
    \centering
    \includegraphics[width=0.94\textwidth]{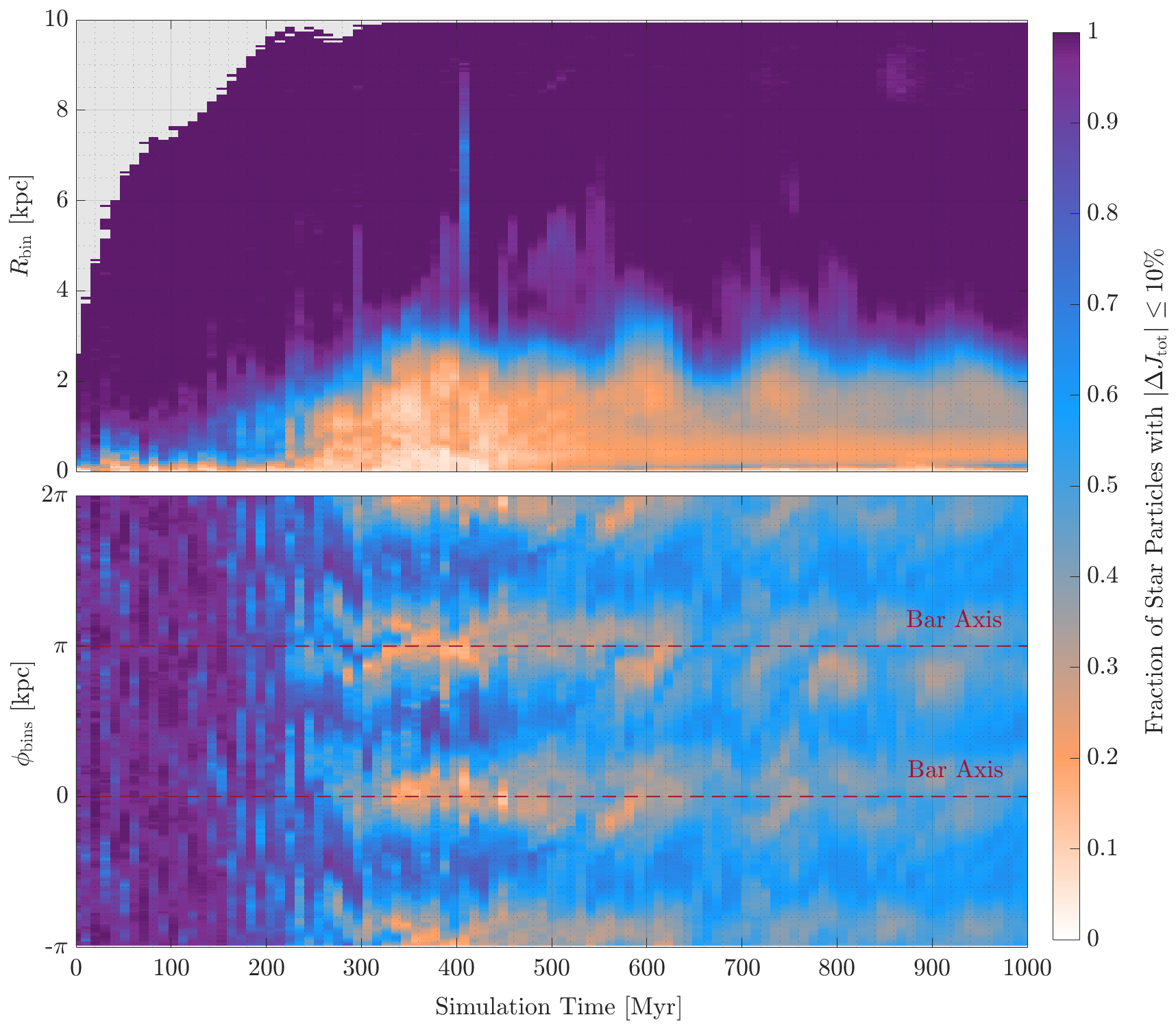}
    \caption{The fraction of stars with total action ($J_{\rm tot}$) changing less than 10\% for each $\Delta t = 10$\,Myr snapshot, separated into radial bins ($500$\,pc, upper) and azimuthal bins ($1.2^\circ$, lower) for the isolated disk (IsoB). The darker the colour the higher the fraction of stars with conserved actions (purple = 100\% conserved, white = 0\%). Horizontal dashed lines indicate the orientation of the bar major axis as described by eye from \citet{iles2025}.}
    \label{f:Jtot_rbint}
\end{figure*}

\begin{figure*}
    \includegraphics[width=0.98\textwidth]{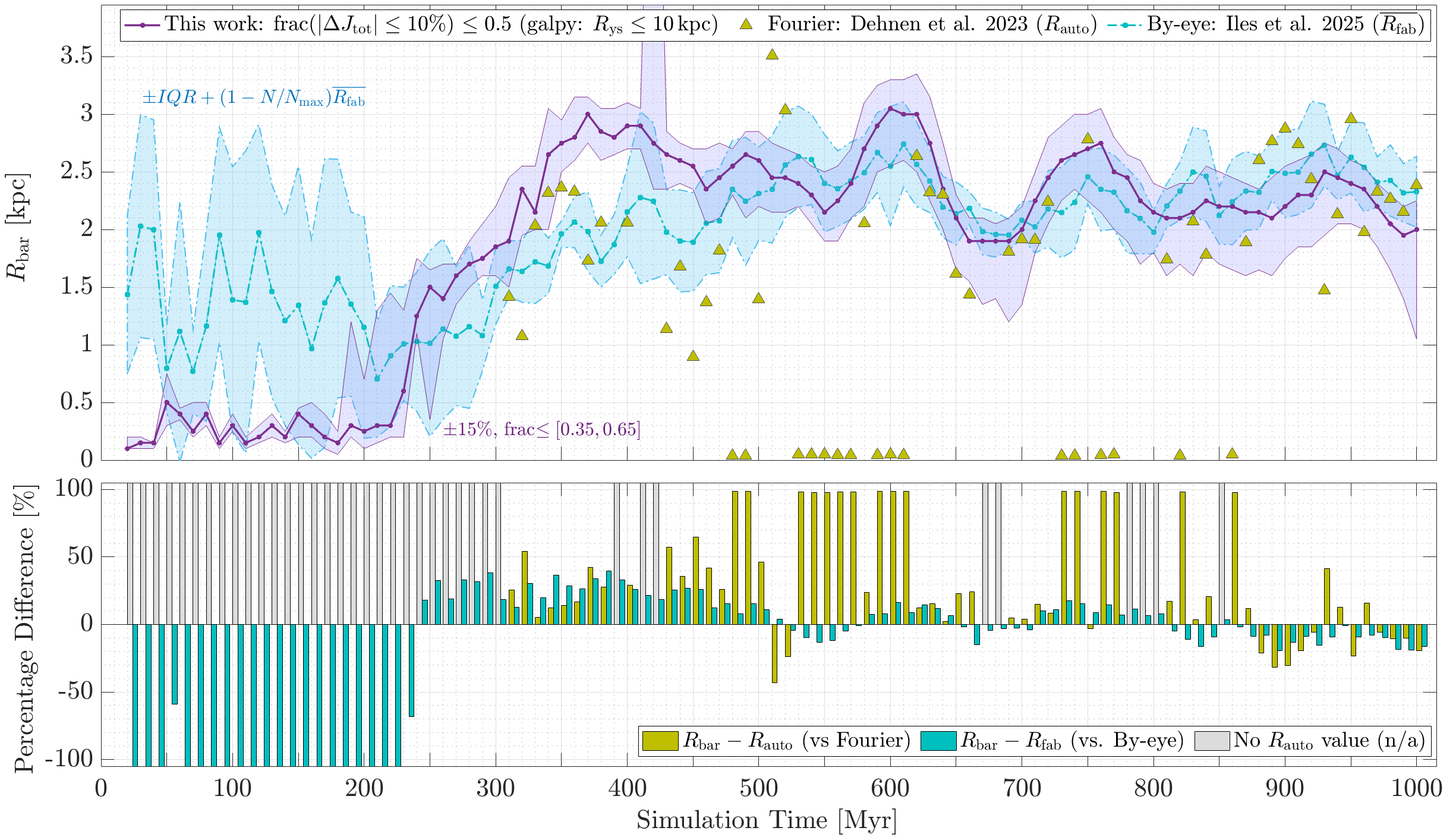}
    \caption{A measure of the bar extent ($R_{\rm bar}$) which is the limit where less than 50\% of star particles in a given radial bin have constant total action ($J_{\rm tot}$ changing by 10\% or less) as calculated by this work (using method A: \textsc{galpy} with simulation-formed stars within $R=10$\,kpc) in purple line, compared with a Fourier bar-finding method \citep{Dehnen2023} in green dashed-line and by-eye \citep{iles2025} in blue dot-dashed-line for the isolated (IsoB) simulation. The bar chart in the lower panel is a representation of the percentage difference between the method employed in this work and the other methods. Grey bars indicate periods where no bar was identified in either of the methods compared.}
    \label{f:Rbar_lit}
\end{figure*}

As a simple representation of how action change per particle correlates with non-axisymmetric structures, Figure\,\ref{f:staractionprj} includes the projected stellar mass density (left) for the $1$\,Gyr snapshot of the isolated (IsoB) simulation, alongside the mean change in total action ($\overline{\Delta J_{\rm tot}}$\,\%, right) calculated along the line-of-sight in grid-square bins (pixels). 
This is to represent the galaxy as if viewed by an external observer, however, we note that the using the full extent in each direction rather than a pixel-width slice could affect the prominence of the structure seen in the image. The resolution of these pixels are each $50$\,pc in $x$ and $y$ and $10$\,pc in $z$ to account for the smaller range in galactic height. The total area considered as part of the galaxy is considered within $R<10$\,kpc and $|z|<1.5$\,kpc as these bounds enclose most of the obvious disk structure and simulated stellar particles. Black contours corresponding to peaks in the mass density profile are also overlaid on the action projection to more easily compare features in this representation. The lighter the colour in the density projection (left) highlights areas of highest density, while more intense colour in the total action (right) denotes areas where the average change in total action is largest. 
The most notable features in the action projection are all centrally located and likely correlate with bar properties. For example, in the edge-on projection, the tip of the X-like feature (or potential boxy-peanut bulge), commonly associated with a barred disk, is significantly more intense than any part of the galaxy outside the very central ($\lesssim0.2$\,kpc) nucleus-like region, where we would expect chaotic orbits. The spokes of the X do not stand out as much as the tips but do appear more evidently green than the surrounding disk contributions, alongside a weak pink-green-pink-green alternating pattern within the disk in the central region ($\lesssim3$\,kpc). In the face-on projection, a similarly intense pink and green alternating pattern is present in the central area, connecting with the edge of the inner high stellar mass density contours. This appears similar to the patterns generated by bar orbits in the centre of galaxies, such as the floret pattern often present in radial velocity maps and associated with the bar. While this is only one snapshot of a single simulated barred galaxy, it validates the hypothesis that the presence of a bar (or strongly non-axisymmetric feature) will affect a correlated and, more importantly, measurable change in the total action of stars (star particles) within these regions in the disk.
Hence, we resolve the first classification criteria for a galaxy's bar: to determine standard parameters describing the bar, such as the radial extent ($R_{\rm bar}$) and orientation angle ($\phi_{\rm bar}$) of the bar within the galaxy. To achieve this classification by using the changes in total action, we prescribe a binary criterion that allows us to separate whether a given star particle has actions which are `conserved' or `not conserved'. We define the action to be functionally conserved if the percentage change in action is less than 10\% of the initial action value. This is a moderate estimate which acts as a buffer for absorbing small systematic errors in the action calculation and simulation process, while preserving most of the true physical variability. We find that choosing a smaller percentage changes the exact fractions of stars with conserved actions, but not the overall behaviour, making this a reasonable threshold.

\begin{figure*}
	\includegraphics[width=0.97\textwidth]{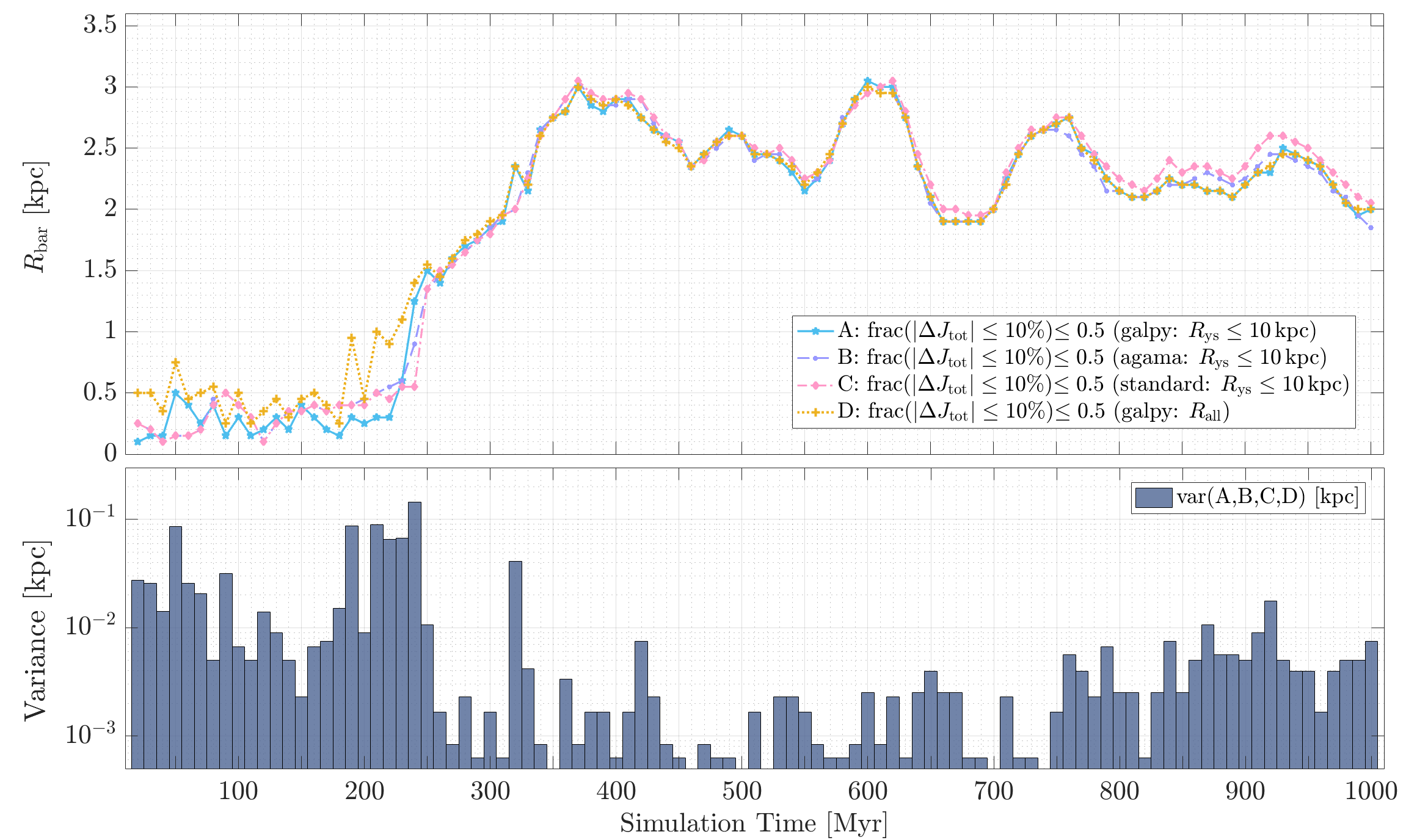}
    \caption{A measure of the bar extent ($R_{\rm bar}$) which is the limit where less than 50\% of star particles in a given radial bin have constant total action ($J_{\rm tot}$ changing by 10\% or less) for the isolated (IsoB) simulation. This are for actions calculated in different methods and populations. A: \textsc{galpy}, B: \textsc{agama}, C: standard MW and D: \textsc{galpy}, with A,B and C constrained to be simulation-formed stars within disk radial extent of $R=10$\,kpc containing most non-axisymmetric features and D containing all possible simulation star particles. The bar chart in the lower panel is a representation of the variance between the A, B, C and D values for each time period.}
    \label{f:Rbar_actions}
\end{figure*}

To determine a measure of the bar extent, we separate the star particles into bins of $50$\,pc in radius, which is equivalent to the stellar softening length in this simulated disk. Similarly, we also separate the star particles into bins of $0.0209$\,radians ($1.2^\circ$) in azimuth to identify the orientation. In each bin, we are able to calculate the fraction of star particles with conserved actions ($|\Delta J_{\rm tot}| \le 10$\%). This is presented for the simulated disk (IsoB) in Figure\,\ref{f:Jtot_rbint}, with radial variation in the upper panel and azimuthal variation in the lower panel. Additionally, the lower panel uses the azimuthal positions from IsoB corrected for disk rotation with the bar along a single axis (aligned with $\phi = 0,\pi$\,rad.) as set from by-eye assessment of this galaxy in \citet[][]{iles2025}. The darker the colour in this figure, the higher the fraction of stars with conserved actions (purple = 100\%, white = 0\% conserved). At first glance, a distinctly light (orange) region can be seen at low radii (upper panel) and mostly aligned with the red dashed lines at $\phi = 0,\pi$\,rad. (lower panel), particularly after approximately $300$\,Myr in simulation time. The radial extent of this region also obviously varies quite frequently, even after the initial growth period ($\sim 200-350$\,Myr), similar to the observed variation of bar extent in many simulations of barred disks, as the instabilities and resonances of the bar settle, the arms disconnect and reconnect, and the galaxy itself evolves. 
Additionally, within this highlighted region, there are two other distinct features. Between $\sim250-550$\,Myr much of the inner galaxy ($\lesssim 2.5$\,kpc) has the lowest recorded fraction of star particles with conserved actions (mostly white coloured bins). This is in the period immediately preceding the barred period \citep[initially determined barred after $\sim 600$\,Myr;][]{iles2022} or precisely during bar formation \citep[bars consistently identified as early as $\sim 300$\,Myr by-eye;][]{iles2025}. This indicates that the largest fraction of stars with non-conserved actions should occur during the initial formation period of the bar and for centrally-located stars in that period. This then settles into a mores stable fraction of 10-50\% of centrally located star particles as the bar evolves in this disk. The second key feature, a thin blue line wrapped by orange/white on either side, with increasing radial-width, is evident in the very inner region (centred on $R\sim 0.25$\,kpc, max. width $\Delta R\sim 0.2$ at $t=1000$\,Myr). This feature becomes apparent after approximately $500$\,Myr and increases in prominence over the next $500$\,Myr of simulation time. This indicates that there is a small feature which must form after the bar and exist only within the inner part of the galaxy, but not the very centre, which increases the fraction of star particles with conserved actions when compared to star particles at surrounding radii. We assert that the conservation of action here could also be tracing the formation of a nuclear stellar disk (NSD). From the kinematic maps (see Appendix \ref{a:nsd} for examples), it does seem that this galaxy also exhibits the standard inner features which would indicate a small NSD, although no specific identification of the potential NSD or its formation time have currently been completed. As the focus of this study is on the bar specifically, we include no further discussion on this topic but simply raise it as a further possible use-case for this method of morphology classification. 

For completeness, we also considered the individual contributions of the action components ($J_R,\,J_\phi,\,J_z$) in projection as Figure\,\ref{f:staractionprj}, as well as in radius and azimuth as in Figure\,\ref{f:Jtot_rbint}. The conservation in each component traces different aspects of the non-axisymmetric structure in the disk as expected, with likely exchange between components over time. The central bar and X-shaped feature are particularly prevalent in $J_R$ and $J_\phi$, while the $J_\phi$ component seems to contribute most to the total shape of the radial distribution in Figure\,\ref{f:Jtot_rbint}. However, categorising how actions are individually affected by galactic structure is beyond the scope of this work. We are confident that the total combination of these components is necessary for classification purposes, as each component contributes to the full response of stars to the presence and impact of the bar.  

To specifically identify a value for the radial bar extent ($R_{\rm bar}$), we set a threshold condition that no more than half of the star particles within the bar radii should be conserved (frac($|\Delta J_{\rm tot}| \le 10$\%)$\le 0.5$), as this non-axisymmetric feature should inherently disrupt action conservation. Subsequently, to identify the bar orientation within this bar extent, we look for two similar height peaks in the fraction of non-conserved star particles ($|\Delta J_{\rm tot}| \le 10$\%) with a separation of approximately $\Delta\phi=\pi$\,rad and record the mean centre of these peaks as the bar orientation angle ($\phi_{\rm bar}$). If a bar semi-minor axis is deemed necessary, it is simply possible to take the full width (FW) or the full-with-at-half-maximum (FWHM) from these peaks in azimuth as an approximate semi-minor axis value. To be more precise, one could also consider using the peak edges as the bar envelope. Although this envelope is difficult to extract and significantly more difficult to implement than a single spatial parameter, we note that there may indeed be some research applications which would benefit from such a detailed method of classification.
To demonstrate the relative consistency of this method against other bar-finding methods, Figure\,\ref{f:Rbar_lit} is included as an example. In this figure, we present a measure of the bar extent ($R_{\rm bar}$) for the simulated disk measured from the non-conserved fraction (frac($|\Delta J_{\rm tot}| \le 10$\%)$\le 0.5$) in a purple solid-line, compared with the by-eye classification of this galaxy from \citet{iles2025}, represented by a blue dot-dashed-line and the publicly available, automated, Fourier-based method of \citet{Dehnen2023} used by the authors of this work on the test galaxy (green triangles, only resolved for snapshots in which the automated method returned a result). This work and the visual classification have coloured regions shaded to give an indication of the confidence interval of these results. As we know that classification by eye is variable, we include the interquartile range (IQR) modified by the number of classifiers confident that a bar exists in a given snapshot \citep[][]{iles2025} as the range of confidence for the by-eye classification. Similarly, the threshold fraction of 50\% was arbitrarily defined so we include a region mapped by moderating this threshold by $\pm 15$\%, that is, bar stars require no more than 35\%  (min) or 65\% (max) of stars have conserved actions. We also include a bar-chart in the lower panel for ease of visual comparison (a larger bar height indicates a larger percentage difference between existing methods and this work; green bars $=$ This Work $-$ Fourier, blue bars $=$ This Work $-$ By-eye, grey bars $=$ no result case). 
Considering this representation, we find that the action method proposed broadly agrees ($\sim 2-25$\% difference after $\sim500$\,Myr) with both the by-eye (blue) and automated Fourier (green) methods, at least when each successfully identifies the bar feature in this galaxy. The period before bar formation in any regime ($0-300$\,Myr) naturally has the least agreement between the methods, which is not unexpected as a bar doesn’t exist in these periods in any case. This is followed by the period between $\sim 300-500$\,Myr where the bar is likely to be actively forming within the disk. In this regime, the differences between methods are larger ($\sim 10-60$\%) which may be indicative of the sensitivity in each method to different tracers of the bar-like feature as it evolves. Since we lack a fundamental definition for the bar or a singular standard method, we also lack a `ground truth' for comparison. Thus, we simply state that this method is capable of reliably reproducing the expected spatial parameters of the bar, with classification power and demonstrated results which are particularly comparable to classifications made by the human eye. This method is also widely applicable at all stages of bar and central‑region development without any additional tuning to the disk conditions and even despite a possible NSD, which is likely to be the cause of difficulty for the automated method to determine the bar extent here (see low $R_{\rm bar}$ values, green dashed-line).

However, we must briefly return to the issue of the galactic potential necessary for determining the action angles. As the algorithm or method used may vary the final axisymmetric potential for calculating the action, and the potential is no necessarily easy to descibe for observed galaxies, we present Figure\,\ref{f:Rbar_actions} to confirm that these potential differences do not, in turn, significantly differ the results of an action-based definition of the bar. This figure is similar to the previous Figure\,\ref{f:Rbar_actions} with the measure of the bar extent ($R_{\rm bar}$) determined from the fraction of star particles at a given radii with conserved total action (frac($|\Delta J_{\rm tot}| \le 10$\%)$\le 0.5$) in the simulated disk. In this figure, the upper panel includes the $R_{\rm bar}$ measured for each different potential (A,D: \textsc{galpy}, B: \textsc{agama}, C: standard MW) and in two different stellar populations, one with only simulation-formed star particles (ys: age$<1000$\,Myr) within disk radial extent of $R=10$\,kpc, as in previous figures, and one including all simulated stars (A,B\&C: ys, $R\le10$\,kpc, D: all stars, all possible radii). Option A is represented by a blue solid-line, B by a purple dashed-line, C by a pink dot-dashed-line, and D by a yellow dotted-line. The bar-chart in the lower panel now serves as a visual representation of the variance between $R_{\rm bar}$ values from options A, B, C and D for each time period and the $y-$axis is $\log\_{10}$(variance). Visually, there is evidently some difference between different potentials, different action calculation methods and very slightly between young and old stellar populations in the simulation. However, the variance between these results is low ($\lesssim0.1$\,kpc), particularly in the period we care most about ($\lesssim 0.01$\,kpc, $t\gtrsim 300$\,Myr) when we are likely tracing a bar feature. We can be confident that regardless of the potential chosen (even a standard form of the potential derived from MW data and scaled to match our simulation), this method is able to recover similar spatial parameters for the bar in this galaxy. 

\begin{figure}
    \centering
	\includegraphics[width=0.78\linewidth]{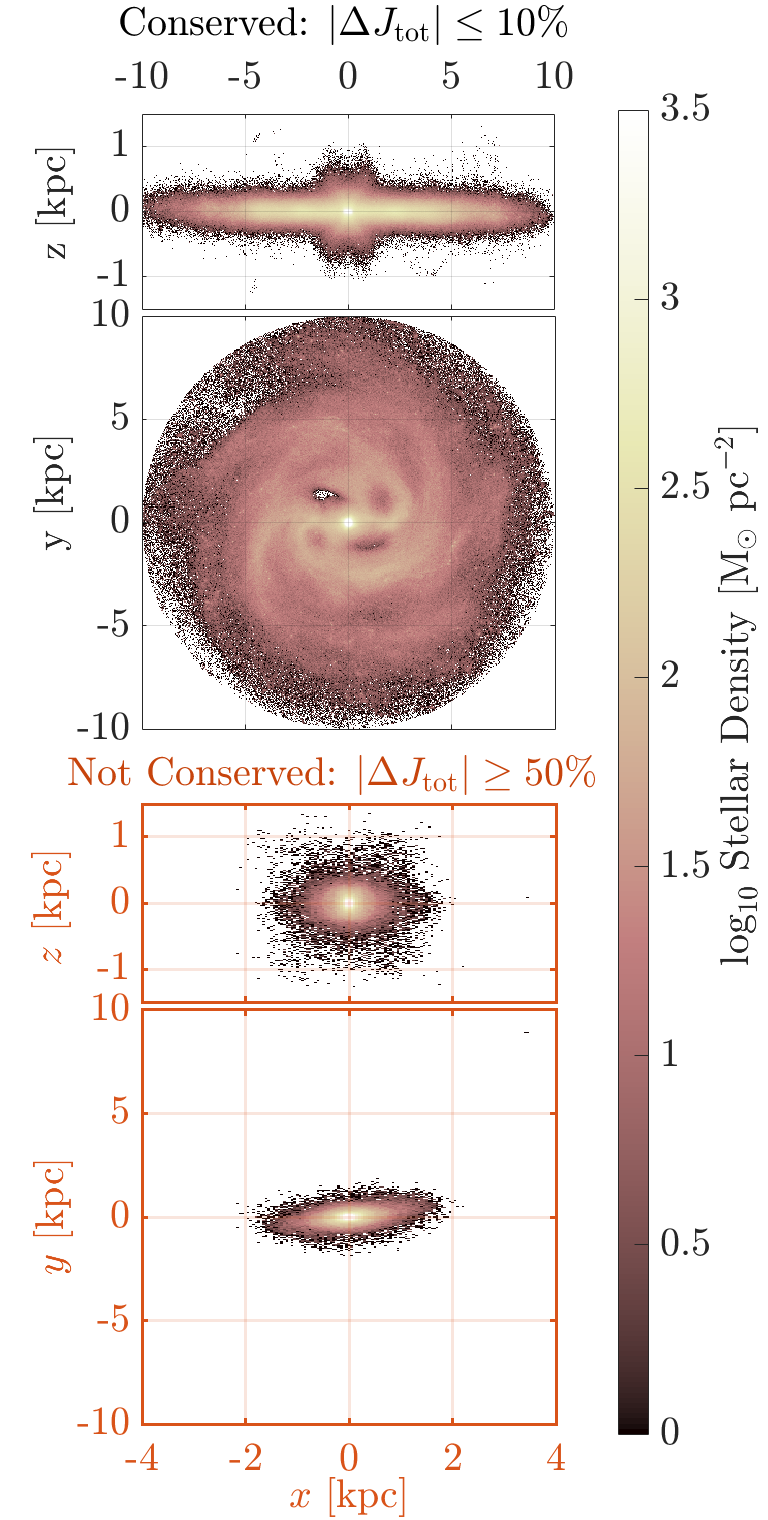}
    \caption{The stellar mass density projected in edge-on ($xz$ - upper) and face-on ($xy$ - lower) orientation for the $1$\,Gyr snapshot, using the same grid-size of $x=y=50$ and $z=10$\,pc as Figure \ref{f:staractionprj}, for the conserved (upper, total action $J_{\rm tot}$ changing by 10\% or less) and definitely not conserved (lower, total action $J_{\rm tot}$ changing by 50\% or more) star particles in the isolated (IsoB) simulation. Note that the not conserved panel is zoomed-in to highlight the features with non-conserved actions (axes labelled in red).}
    \label{f:action_barstars}
\end{figure}

Finally, we note that this method was also tested on the tidally-driven disk (TideB) from \citet{iles2022}, which is undergoing an $S=10$\% interaction with a small stellar companion similar to a dwarf galaxy. This interaction perturbs many properties and features in the main disk while it also promotes the formation of the bar. If dealing with non-isolated galaxy conditions, it is important to ensure that the main disk is properly centred before applying any method of action calculation. If the disk is incorrectly centred, the implementation of the potential and the subsequent actions returned, including the conservation of these actions, will be affected and action conservation cannot be used to identify specific features. However, if the main disk can be properly located in the potential, regardless of any interacting companions, we find no reason that this method not be applicable to galaxies in complex interacting systems. More care must simply be taken in the initial setup and action calculation. 

In general, this test demonstrates that the conservation of action method described is able to perform at least as well, if not better, than existing automated methods and visual classification for identifying the barred region in galaxies, even with complex inner morphologies (e.g. a NSD) or complex environments (e.g. small/early-stage interactions or mergers).

%%%%%%%%%%%%%%%%%%%%%%%%%%%%%%%%%%%%%%%%%%%%%%%%%%%%%%%%%%
\section{Star-by-Star Classification}
\label{s:starbystar}

\begin{figure}
    \centering
	\includegraphics[width=0.99\linewidth]{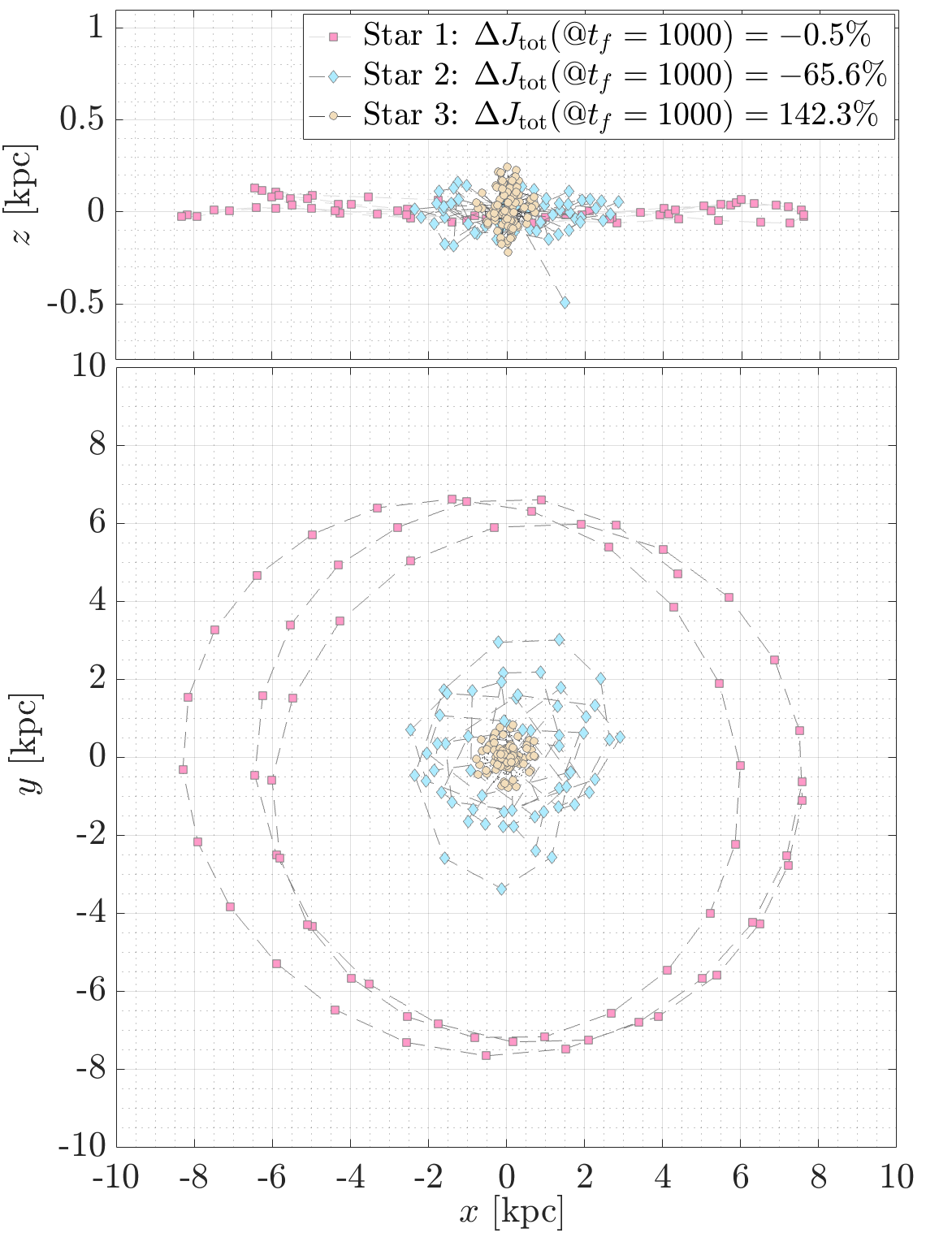}
    \caption{Example orbits for star particles with a change in total action $J_{\rm tot} \sim 0$\% (pink squares), $\sim 50$\% (blue diamonds) and $>100$\% (yellow circles) respectively, as measured for the particles at the 1\,Gyr snapshot (on $\Delta t = 10$\,Myr). These orbits have not been corrected for bar precession (i.e. not presented in the bar reference frame).}
    \label{f:actionstars}
\end{figure}

The benefit of using the conservation of action to classify the bar is that the actions must be first calculated on a star-by-star basis. This means that it is possible to identify whether a given star particle in the simulation is likely to be on a bar orbit, helping us to exclude star particles which may simply be passing through the identified bar region. This kind of star-by-star classification could then allow for more precise studies of the stellar populations or properties of the bar, rather than simply of bar-located stellar populations or properties. An example of this is presented in Figure\,\ref{f:action_barstars}. In this figure, the stellar mass density is again projected edge-on ($xz$ - upper) and face-on ($xy$ - lower) orientations but the two columns separate populations of star particles with clearly conserved and not conserved total actions (upper - conserved $|\Delta J_{\rm tot}|\le 10$\%, lower, red outline - not conserved $|\Delta J_{\rm tot}|\ge 50$\%) as measured for the 1\,Gyr final snapshot of the IsoB simulation (on $\Delta t = 10$\,Myr). The region of the galaxy mapped by the star particles with non-conserved actions is significantly smaller, as these predominantly occupy the central region of the galaxy, so we have enhanced the axis limits in the lower panels of this figure (highlighted in red) to similarly focus on the features in each.
While a bar-like over-density feature still exists in the face-on projection and the X or boxy-peanut shape is still evident edge-on projection of the `conserved' population (upper panels), this has much to do with the moderate $10$\% range we use to consider conservation. Reducing this percentage does indeed reduce the contamination of bar-like features in these projections. In comparison, the `not conserved' population is represented face-on by precisely a long ellipsoid-like feature and a square-like boxy-peanut shape with extended X-like tips in the edge-on projection (red, lower panels) which is exactly what we would expect from a bar. 

As the intent of this work is to demonstrate the functionality of action for separating the bar from the surrounding galaxy, we first demonstrate projection, as in Figure\,\ref{f:action_barstars}, but are more focussed on demonstrating the classification on a star-by-star basis. It is inherently possible for this method to identify any individual star particle in the simulation with different levels of change in action, and to classify these respectively. In Figure\,\ref{f:actionstars}, we select one star with each: the change in total action $J_{\rm tot} \sim 0$\% (pink), $\sim 50$\% (blue) and $>100$\% (yellow) respectively, as measured for the particles at the 1\,Gyr snapshot (on $\Delta t = 10$\,Myr), and plot orbits throughout the lifetime of each star particle. These three stars are indicative of each: a mostly circular disk orbit ($R\gtrsim5$\,kpc), a bar-orbit ($3\gtrsim R\gtrsim 0.5$\,kpc), and a central orbit ($R\lesssim0.5$\,kpc). These star particles were randomly selected based purely on their change in total action percentage ($\Delta J_{\rm tot}$\%) as a means to demonstrate the additional power of this method for broadly separating orbits that commonly inhabit different spatial/morphological regions of the galactic disk, without having to analyse the full orbits of each star. As such, with this method it is possible to define the bar region in projection via standard bar parameters, while similarly exploiting the capability to separate the bar-located and bar-member stellar populations based on the star-by-star values of the total change in action. This is practical for simulation studies where all stellar particles are resolved and change can be measured directly from snapshot-to-snapshot time evolution. However, as we demonstrate in \citep[][]{iles2026} this concept can also be exploited in observations where actions are similarly calculated on a star-by-star basis, such as within the Milky Way.

%%%%%%%%%%%%%%%%%%%%%%%%%%%%%%%%%%%%%%%%%%%%%%%%%%%%%%%%%%%%%%%%%%%
\section{Conclusions}
\label{s:conclusion}

Many methods exist for classifying and identifying the bar-like features in the morphology of many disk-type galaxies. Observations and simulations each present a different set of challenges and opportunities for defining what precisely constitutes the bar in the centre of a galaxy and consequently, how this feature may drive, impact and effect the surrounding galactic structure and properties. Unfortunately, until one distinct and unified method is developed, which is universally applicable, we will always struggle to ensure like-with-like comparisons across studies using different methods to define the bar. This, in turn, has lasting implications for our understanding of galaxy formation and evolution in areas influenced by the bar. 
Here, we demonstrate through a simulated test galaxy \citep[IsoB;][]{iles2022} that the conservation of action may be a viable method to determine the spatial parameters (e.g. radial extent $R_{\rm bar}$, orientation $\phi_{\rm bar}$) often used to describe the bar region in galaxies. It is also possible to use this property to identify stars (or star particles in a simulation) which are more likely to be on bar orbits and thus, identifying members of the bar on a star-by-star basis. 

By exploiting the condition of axisymmetry necessary for the action integrals to be considered adiabatic invariant quantities, we use the conservation of action on a short timescale ($\Delta t = 10$\,Myr) for each star (star particle) as a metric for whether a star is likely to be on any quasi-periodic orbit within an axisymetric potential or, instead, located in a strongly non-axisymmetric part of the galactic potential (i.e. the bar). Setting a threshold of $<10$\% change in value of the total action ($\Delta J_{\rm tot}$) over the time period to be considered conserved actions, we can determine the fraction of stars throughout the disk with conserved or not conserved action. By using a fraction of no more than $0.5$ (or $50$\%) of star particles with conserved actions ($|\Delta J_{\rm tot}|<10$\%) in a given spatial bin as a limit for a strongly non-axisymmetric structure, we are able to recover bar classification parameters, such as the bar length ($R_{\rm bar}$) and orientation ($\phi_{\rm bar}$). For the test simulation, the action conservation calculation was achieved for every snapshot and the returned values similar in magnitude and variation across the evolution of the disk to by-eye classifications of the same galaxy. These values are also similar in magnitude to results from an automated, Fourier-based method of bar classification for this galaxy where these are returned, but appears more resilient to complex central structures (e.g. NSD formation) and bar-arm decoupling, as well as 
noisy systems, such as 
gas-rich stellar discs with a highly turbulent ISM \citep{Bland-Hawthorn2024,Bland-Hawthorn2025,Zhang2025}.
As there is currently no available `true' solution to test against the actual bar parameters, we propose that this method appears to be a robust and viable means to identify the bar in galaxies where the action is calculable for a large percentage of stars (star particles) and where it is possible to measure the change in action over time ($\Delta t \le 100$\,Myr). 

In addition to standard bar region classification, the change in action ($\Delta J_{\rm tot}$\%) can also be used on a star-by-star basis to assess the bar membership of stars from an orbital perspective, as action is fundamentally related to orbit dynamics. Star particles with low changes in total action ($|\Delta J_{\rm tot}| \sim 0$\%) are found to be mainly located at larger disk radii and on predominantly circular orbits. Star particles with moderate changes in total action ($|\Delta J_{\rm tot}| \sim 50$\%) are found to be mostly located at inner radii, within co-rotation and the spatially determined bar extent, on more complex orbits (e.g. in bar orbit families). Finally, star particles with the largest changes in total action ($|\Delta J_{\rm tot}| \gtrsim 100$\%) are almost all centrally located ($R<1$\,kpc) with primarily chaotic orbits. In this way, the change in action may also allow for a distinction between stars in the bar region but not following bar-like orbits (passing through) compared to the true members of the bar (on bar orbits). This has the potential to impact ongoing studies of bar-related galactic properties, such as star formation, stellar migration and chemical mixing. 

We recommend the conservation of action as an alternative, repeatable, method for classifying bars, at least in simulations of galaxies. Through this work we have sought to demonstrate the following. 
\vspace{-0.15cm}
\begin{itemize}
\item{It is possible to use the conservation of action to classify the spatial extent of barred region in simulated galaxies.}
\vspace{0.05cm}
\item{The percentage change in action can be a good indicator for stellar orbits that contribute to different morphological features in simulated galaxies. }
\end{itemize}
\vspace{-0.15cm}
While this method, too, has both limitations and advantages, we hope that this work serves to advance ongoing efforts within the field to unify the classification of bars in galaxies and supports future research into bars, the galaxies that host them, and the evolution of structure in the Universe on the whole. 

%%%%%%%%%%%%%%%%%%%%%%%%%%%%%%%%%%%%%%%%%%%%%%%%%%
\section*{Acknowledgements}
This work employed the \textsc{python} package of \textsc{pynbody} \citep[][\url{https://github.com/pynbody/pynbody}]{Pontzen2013} and the \textsc{matlab} 
platform \citep{matlab}, as well as the Astrophysics Data System, funded by NASA under Cooperative Agreement 80NSSC21M00561. EJI acknowledges the support of the Australian Research Council through Discovery Project DP220103384 (CIs:\,JBH,\,KF). FP acknowledges support from the Australian Government Research Training Program Scholarship and from the UNSW Scientia Program. We thank the anonymous referee for their constructive comments.

%%%%%%%%%%%%%%%%%%%%%%%%%%%%%%%%%%%%%%%%%%%%%%%%%%
\section*{Data Availability}
The data underlying this article can be shared upon reasonable request to the corresponding author.

%%%%%%%%%%%%%%%%%%%% REFERENCES %%%%%%%%%%%%%%%%%%

\bibliographystyle{mnras}
\bibliography{AAmybib}

@ARTICLE{Zhang2025,
       author = {{Zhang}, HanYuan and {Tepper-Garc{\'\i}a}, Thor and {Belokurov}, Vasily and {Evans}, N. Wyn and {Tsukui}, Takafumi and {Davis}, Hillary and {Bland-Hawthorn}, Joss and {Sanders}, Jason L. and {Agertz}, Oscar},
        title = "{Enhanced rates of stellar radial migration in gas-rich discs at high redshift}",
      journal = {arXiv e-prints},
         year = 2025,
        month = dec,
          eid = {arXiv:2512.09030},
        pages = {arXiv:2512.09030},
          doi = {10.48550/arXiv.2512.09030},
archivePrefix = {arXiv},
       eprint = {2512.09030},
 primaryClass = {astro-ph.GA},
       adsurl = {https://ui.adsabs.harvard.edu/abs/2025arXiv251209030Z}
}

@ARTICLE{Bland-Hawthorn2025,
       author = {{Bland-Hawthorn}, Joss and {Tepper-Garcia}, Thor and {Agertz}, Oscar and {Federrath}, Christoph and {Haywood}, Misha and {di Matteo}, Paola and {Bedding}, Timothy R. and {Tsukui}, Takafumi and {Wisnioski}, Emily and {Ness}, Melissa and {Freeman}, Ken},
        title = "{Turbulent Gas-rich Disks at High Redshift: Origin of Thick Stellar Disks Through 3D ``Baryon Sloshing''}",
      journal = {\apj},
         year = 2025,
        month = nov,
       volume = {994},
       number = {1},
          eid = {22},
        pages = {22},
          doi = {10.3847/1538-4357/ae0931},
archivePrefix = {arXiv},
       eprint = {2502.01895},
 primaryClass = {astro-ph.GA},
       adsurl = {https://ui.adsabs.harvard.edu/abs/2025ApJ...994...22B}
}

@ARTICLE{iles2022,
author = {{Iles}, Elizabeth J. and {Pettitt}, Alex R. and {Okamoto}, Takashi},
title = "{Differences in star formation activity between tidally triggered and isolated bars: a case study of NGC 4303 and NGC 3627}",
journal = {\mnras},
year = 2022,
month = mar,
volume = {510},
number = {3},
pages = {3899-3916},
doi = {10.1093/mnras/stab3330},
}

@ARTICLE{post15,
       author = {{Posti}, Lorenzo and {Binney}, James and {Nipoti}, Carlo and {Ciotti}, Luca},
        title = "{Action-based distribution functions for spheroidal galaxy components}",
      journal = {\mnras},
         year = 2015,
        month = mar,
       volume = {447},
       number = {4},
        pages = {3060-3068},
          doi = {10.1093/mnras/stu2608},
archivePrefix = {arXiv},
       eprint = {1411.7897},
 primaryClass = {astro-ph.GA},
       adsurl = {https://ui.adsabs.harvard.edu/abs/2015MNRAS.447.3060P}
}

@ARTICLE{will14,
       author = {{Williams}, A.~A. and {Evans}, N.~W. and {Bowden}, A.~D.},
        title = "{Hamiltonians of spherical Galaxies in action-angle coordinates}",
      journal = {\mnras},
         year = 2014,
        month = aug,
       volume = {442},
       number = {2},
        pages = {1405-1410},
          doi = {10.1093/mnras/stu892},
archivePrefix = {arXiv},
       eprint = {1405.1065},
 primaryClass = {astro-ph.GA},
       adsurl = {https://ui.adsabs.harvard.edu/abs/2014MNRAS.442.1405W}
}

@ARTICLE{iles2024,
author = {{Iles}, Elizabeth J. and {Pettitt}, Alex R. and {Okamoto}, Takashi and {Kawata}, Daisuke},
title = "{The impact of bar origin and morphology on stellar migration}",
journal = {\mnras},
year = 2024,
month = jan,
volume = {527},
number = {2},
pages = {2799-2815},
doi = {10.1093/mnras/stad3377},
}

@ARTICLE{iles2025,
author = {{Iles}, Elizabeth J. and Joss {Bland-Hawthorn} and Courtney {Crawford} and Scott {Croom} and Hillary {Davis} and May Gade {Pedersen} and Anne {Green} and Madusha {Gunawardhana} and Miguel {Icaza-Lizaola} and Helen {Johnston} and Emily F. {Kerrison} and Yifan {Mai} and Benjamin T. {Montet} and Kovi {Rose} and Tomas {Rutherford} and Manasvee {Saraf} and Ellen L. {Sirks} and Eckhart {Spalding} and Sujeeporn {Tuntipong} and Jesse {van de Sande} and Pavadol {Yamsiri}},
title = "{findAbar: How Astronomers may Perceive the Bar in Galaxies Differently.}",
journal = {\pasa},
year = 2025,
volume = {42},
pages = {e166},
doi = {10.1017/pasa.2025.10126}
}

@article{iles2026,
author = {{Iles}, Elizabeth J. and Finn A. {Pal} and Joss {Bland-Hawthorn} and Ken {Freeman} and Sarah {Martell}},
title = "{Finding the Milky Way Bar via Action Conservation in Stellar Surveys}",
journal = {\mnras},
year = 2026,
month = jan,
volume = {in prep.},
}

@article{Abraham1999,
author = {Abraham, R. G. and Merrifield, M. R. and Ellis, R. S. and Tanvir, N. R. and Brinchmann, J.},
doi = {10.1046/j.1365-8711.1999.02766.x},
journal = {\mnras},
month = {sep},
number = {2},
pages = {569--576},
title = {{The evolution of barred spiral galaxies in the Hubble Deep Fields North and South}},
volume = {308},
year = {1999}
}

@article{Abraham2018,
author = {Abraham, Sheelu and Aniyan, A K and Kembhavi, Ajit K and Philip, N S and Vaghmare, Kaustubh},
doi = {10.1093/mnras/sty627},
journal = {\mnras},
month = {jun},
number = {1},
pages = {894--903},
title = {{Detection of bars in galaxies using a deep convolutional neural network}},
volume = {477},
year = {2018}
}

@ARTICLE{Arunima2025,
author = {{Arunima}, Arunima and {Krumholz}, Mark R. and {Ireland}, Michael J. and {Zhang}, Chuhan and {Hu}, Zipeng},
title = "{Fundamental limits to orbit reconstruction due to non-conservation of stellar actions in a Milky Way-like simulation}",
journal = {\mnras},
year = 2025,
month = oct,
volume = {543},
number = {1},
pages = {358-374},
doi = {10.1093/mnras/staf1515},
}

@article{Aguerri1998,
author = {Aguerri, J. A. L. and Beckman, J. E. and Prieto, M.},
doi = {10.1086/300615},
journal = {\aj},
month = {nov},
number = {5},
pages = {2136--2153},
title = {{Bar Strengths, Bar Lengths, and Corotation Radii, Derived Photometrically for 10 Barred Galaxies}},
volume = {116},
year = {1998}
}

@article{Aguerri2009,
author = {Aguerri, J. A. L. and M{\'{e}}ndez-Abreu, J. and Corsini, E. M.},
doi = {10.1051/0004-6361:200810931},
journal = {\aap},
month = {feb},
number = {2},
pages = {491--504},
title = {{The population of barred galaxies in the local universe}},
volume = {495},
year = {2009}
}

@article{Alonso2014,
author = {Alonso, Sol and Coldwell, Georgina and Lambas, Diego G.},
doi = {10.1051/0004-6361/201424523},
journal = {\aap},
month = {dec},
pages = {A86},
title = {{AGN spiral galaxies in groups: effects of bars}},
volume = {572},
year = {2014}
}

@article{Antoja2018,
author = {Antoja, T. and Helmi, A. and Romero-G{\'{o}}mez, M. and Katz, D. and Babusiaux, C. and Drimmel, R. and Evans, D. W. and Figueras, F. and Poggio, E. and Reyl{\'{e}}, C. and Robin, A. C. and Seabroke, G. and Soubiran, C.},
doi = {10.1038/s41586-018-0510-7},
journal = {Nature},
month = {sep},
number = {7723},
pages = {360--362},
title = {{A dynamically young and perturbed Milky Way disk}},
volume = {561},
year = {2018}
}

@book{Arnold1978,
author = {{Arnold}, Vladimir Igorevich},
title = "{Graduate texts in mathematics: Mathematical methods of classical mechanics}",
year = 1978,
publisher = {New York: Springer},
}

@article{Athanassoula2002,
author = {Athanassoula, E.},
doi = {10.1086/340784},
journal = {\apj},
month = {apr},
number = {2},
pages = {L83--L86},
title = {{Bar-Halo Interaction and Bar Growth}},
volume = {569},
year = {2002}
}

@article{Athanassoula2005,
author = {Athanassoula, E.},
doi = {10.1007/s10569-004-4947-7},
journal = {Celestial Mechanics and Dynamical Astronomy},
month = {jan},
number = {1-2},
pages = {9--31},
title = {{Dynamical Evolution of Barred Galaxies}},
volume = {91},
year = {2005}
}

@article{Athanassoula2013,
author = {Athanassoula, E. and Machado, Rubens E. G. and Rodionov, S. A.},
doi = {10.1093/mnras/sts452},
journal = {\mnras},
month = {mar},
number = {3},
pages = {1949--1969},
title = {{Bar formation and evolution in disc galaxies with gas and a triaxial halo: morphology, bar strength and halo properties}},
volume = {429},
year = {2013}
}

@article{Barbanis1967,
author = {Barbanis, B. and Woltjer, L.},
doi = {10.1086/149349},
journal = {The Astrophysical Journal},
month = {nov},
pages = {461},
title = {{Orbits in Spiral Galaxies and the Velocity Dispersion of Population i Stars}},
volume = {150},
year = {1967}
}

@BOOK{Binney2008,
author = {{Binney}, James and {Tremaine}, Scott},
title = "{Galactic Dynamics: Second Edition}",
year = 2008,
publisher = {Princeton Univ. Press, Princeton}
}

@article{Binney2012,
author = {Binney, James},
doi = {10.1111/j.1365-2966.2012.21757.x},
issn = {00358711},
journal = {Monthly Notices of the Royal Astronomical Society},
month = {oct},
number = {2},
pages = {1324--1327},
title = {{Actions for axisymmetric potentials}},
volume = {426},
year = {2012}
}

@article{Binney2020,
author = {Binney, James},
doi = {10.1093/mnras/staa092},
journal = {Monthly Notices of the Royal Astronomical Society},
month = {jun},
number = {1},
pages = {886--894},
title = {{Angle-action variables for orbits trapped at a Lindblad resonance}},
volume = {495},
year = {2020}
}

@ARTICLE{Bland-Hawthorn2016,
author = {{Bland-Hawthorn}, Joss and {Gerhard}, Ortwin},
title = "{The Galaxy in Context: Structural, Kinematic, and Integrated Properties}",
journal = {\araa},
year = 2016,
month = sep,
volume = {54},
pages = {529-596},
doi = {10.1146/annurev-astro-081915-023441},
archivePrefix = {arXiv},
eprint = {1602.07702}
}

@ARTICLE{Bland-Hawthorn2024,
       author = {{Bland-Hawthorn}, Joss and {Tepper-Garcia}, Thor and {Agertz}, Oscar and {Federrath}, Christoph},
        title = "{Turbulent Gas-rich Disks at High Redshift: Bars and Bulges in a Radial Shear Flow}",
      journal = {\apj},
         year = 2024,
        month = jun,
       volume = {968},
       number = {2},
          eid = {86},
        pages = {86},
          doi = {10.3847/1538-4357/ad4118},
archivePrefix = {arXiv},
       eprint = {2402.06060},
 primaryClass = {astro-ph.GA},
       adsurl = {https://ui.adsabs.harvard.edu/abs/2024ApJ...968...86B}
}

@article{Blitz1991,
author = {Blitz, Leo and Spergel, David N.},
doi = {10.1086/170535},
journal = {\apj},
month = {oct},
pages = {631},
title = {{Direct evidence for a bar at the Galactic center}},
volume = {379},
year = {1991}
}

@ARTICLE{Bovy2015,
author = {{Bovy}, Jo},
title = "{galpy: A python Library for Galactic Dynamics}",
journal = {\apjs},
year = 2015,
month = feb,
volume = {216},
number = {2},
eid = {29},
pages = {29},
doi = {10.1088/0067-0049/216/2/29},
}

@article{Carlberg1985,
author = {Carlberg, R. G. and Sellwood, J. A.},
doi = {10.1086/163134},
journal = {The Astrophysical Journal},
month = {may},
pages = {79},
title = {{Dynamical evolution in galactic disks}},
volume = {292},
year = {1985}
}

@ARTICLE{Carlberg2024,
author = {{Carlberg}, Raymond G. and {Jenkins}, Adrian and {Frenk}, Carlos S. and {Cooper}, Andrew P.},
title = "{Star Stream Velocity Distributions in Cold Dark Matter and Warm Dark Matter Galactic Halos}",
journal = {\apj},
year = 2024,
month = nov,
volume = {975},
number = {1},
pages = {135},
doi = {10.3847/1538-4357/ad7b35},
}

@article{Casasola2011,
author = {Casasola, V. and Hunt, L. K. and Combes, F. and Garc{\'{i}}a-Burillo, S. and Neri, R.},
doi = {10.1051/0004-6361/201015680},
issn = {00046361},
journal = {Astronomy and Astrophysics},
number = {11},
title = {{Molecular gas in NUclei of GAlaxies (NUGA): XIV. the barred LINER/Seyfert 2 galaxy NGC3627}},
volume = {527},
year = {2011}
}

@article{Cavanagh2020,
author = {Cavanagh, M. K. and Bekki, K.},
doi = {10.1051/0004-6361/202037963},
journal = {\aap},
month = {sep},
pages = {A77},
title = {{Bars formed in galaxy merging and their classification with deep learning}},
volume = {641},
year = {2020}
}

@article{Cavanagh2022,
author = {Cavanagh, Mitchell K and Bekki, Kenji and Groves, Brent A and Pfeffer, Joel},
doi = {10.1093/mnras/stab3786},
journal = {\mnras},
month = {jan},
number = {4},
pages = {5164--5178},
title = {{The evolution of barred galaxies in the EAGLE simulations}},
volume = {510},
year = {2022}
}

@article{Cavanagh2024,
author = {Cavanagh, Mitchell K and Bekki, Kenji and Groves, Brent A},
doi = {10.1093/mnras/stae801},
issn = {0035-8711},
journal = {\mnras},
month = {apr},
number = {1},
pages = {1171--1194},
title = {{A morphological segmentation approach to determining bar lengths}},
volume = {530},
year = {2024}
}

@article{Chabrier2003,
author = {Chabrier, Gilles},
doi = {10.1086/376392},
journal = {Publications of the Astronomical Society of the Pacific},
month = {jul},
number = {809},
pages = {763--795},
title = {{Galactic Stellar and Substellar Initial Mass Function}},
volume = {115},
year = {2003}
}

@article{Combes1993,
author = {Combes, F. and Elmegreen, B. G.},
journal = {\aap},
pages = {391--401},
title = {{Bars in early- and late-type galaxies}},
volume = {271},
year = {1993}
}

@article{Consolandi2016,
author = {Consolandi, G.},
doi = {10.1051/0004-6361/201629115},
journal = {\aap},
month = {nov},
pages = {A67},
title = {{Automated bar detection in local disk galaxies from the SDSS}},
volume = {595},
year = {2016}
}

@ARTICLE{Costantin2023,
author = {{Costantin}, Luca and {P{\'e}rez-Gonz{\'a}lez}, Pablo G. and {Guo}, Yuchen and {Buttitta}, Chiara and {Jogee}, Shardha and {Bagley}, Micaela B. and {Barro}, Guillermo and {Kartaltepe}, Jeyhan S. and {Koekemoer}, Anton M. and {Cabello}, Cristina and {Corsini}, Enrico Maria and {M{\'e}ndez-Abreu}, Jairo and {de la Vega}, Alexander and {Iyer}, Kartheik G. and {Bisigello}, Laura and {Cheng}, Yingjie and {Morelli}, Lorenzo and {Arrabal Haro}, Pablo and {Buitrago}, Fernando and {Cooper}, M.~C. and {Dekel}, Avishai and {Dickinson}, Mark and {Finkelstein}, Steven L. and {Giavalisco}, Mauro and {Holwerda}, Benne W. and {Huertas-Company}, Marc and {Lucas}, Ray A. and {Papovich}, Casey and {Pirzkal}, Nor and {Seill{\'e}}, Lise-Marie and {Vega-Ferrero}, Jes{\'u}s and {Wuyts}, Stijn and {Yung}, L.~Y. Aaron},
title = "{A Milky Way-like barred spiral galaxy at a redshift of 3}",
journal = {\nat},
year = 2023,
month = nov,
volume = {623},
number = {7987},
pages = {499-501},
doi = {10.1038/s41586-023-06636-x},
}

@article{Daniel2015,
author = {Daniel, Kathryne J. and Wyse, Rosemary F. G.},
doi = {10.1093/mnras/stu2683},
journal = {Monthly Notices of the Royal Astronomical Society},
month = {mar},
number = {4},
pages = {3576--3592},
title = {{Constraints on radial migration in spiral galaxies – I. Analytic criterion for capture at corotation}},
volume = {447},
year = {2015}
}

@article{Debattista2020,
author = {Debattista, Victor P and Liddicott, David J and Khachaturyants, Tigran and {Beraldo e Silva}, Leandro},
doi = {10.1093/mnras/staa2568},
journal = {Monthly Notices of the Royal Astronomical Society},
month = {sep},
number = {3},
pages = {3334--3350},
title = {{Box/peanut-shaped bulges in action space}},
volume = {498},
year = {2020}
}

@ARTICLE{Dehnen1993,
author = {{Dehnen}, W.},
title = "{A Family of Potential-Density Pairs for Spherical Galaxies and Bulges}",
journal = {\mnras},
year = 1993,
month = nov,
volume = {265},
pages = {250},
doi = {10.1093/mnras/265.1.250}
}

@article{Dehnen2023,
author = {Dehnen, Walter and Semczuk, Marcin and Sch{\"{o}}nrich, Ralph},
doi = {10.1093/mnras/stac3184},
journal = {\mnras},
month = {nov},
number = {2},
pages = {2712--2718},
title = {{Measuring bar pattern speeds from single simulation snapshots}},
volume = {518},
year = {2023}
}

@article{Diaz-Garcia2016,
author = {D{\'{i}}az-Garc{\'{i}}a, S. and Salo, H. and Laurikainen, E. and Herrera-Endoqui, M.},
doi = {10.1051/0004-6361/201526161},
journal = {\aap},
month = {mar},
pages = {A160},
title = {{Characterization of galactic bars from 3.6 $\mu$ m S 4 G imaging}},
volume = {587},
year = {2016}
}

@article{Dobbs2006,
author = {Dobbs, C. L. and Bonnell, I. A. and Pringle, J. E.},
doi = {10.1111/j.1365-2966.2006.10794.x},
issn = {0035-8711},
journal = {Monthly Notices of the Royal Astronomical Society},
month = {oct},
number = {4},
pages = {1663--1674},
title = {{The formation of molecular clouds in spiral galaxies}},
volume = {371},
year = {2006}
}

@article{Dobbs2014,
author = {Dobbs, C. L. and Krumholz, M. R. and Ballesteros-Paredes, J. and Bolatto, A. D. and Fukui, Y. and Heyer, M. and {Mac Low}, M.-M. and Ostriker, E. C. and V{\'{a}}zquez-Semadeni, E.},
doi = {10.2458/azu_uapress_9780816531240-ch001},
eprint = {1312.3223},
journal = {Protostars and Planets VI},
title = {{Formation of Molecular Clouds and Global Conditions for Star Formation}},
volume = {3},
year = {2014}
}

@article{Downes1996,
author = {Downes, D. and Reynaud, D. and Solomon, P. M. and Radford, S. J. E.},
doi = {10.1086/177046},
journal = {\apj},
month = {apr},
pages = {186},
title = {{CO in the Barred Galaxy NGC 1530}},
volume = {461},
year = {1996}
}

@article{Drimmel2023,
author = {Drimmel, R. and Khanna, S. and D'Onghia, E. and Tepper-Garc{\'{i}}a, T. and Bland-Hawthorn, J. and Chemin, L. and Ripepi, V. and Romero-G{\'{o}}mez, M. and Ramos, P. and Poggio, E. and Andrae, R. and Blomme, R. and Cantat-Gaudin, T. and Castro-Ginard, A. and Clementini, G. and Figueras, F. and Fouesneau, M. and Fr{\'{e}}mat, Y. and Lobel, A. and Marshall, D. and Muraveva, T.},
doi = {10.1051/0004-6361/202244605},
journal = {Astronomy {\&} Astrophysics},
month = {feb},
pages = {A10},
title = {{A new resonance-like feature in the outer disc of the Milky Way}},
volume = {670},
year = {2023}
}

@article{Durbala2008,
author = {Durbala, A. and Sulentic, J. W. and Buta, R. and Verdes-Montenegro, L.},
doi = {10.1111/j.1365-2966.2008.13713.x},
journal = {\mnras},
month = {nov},
number = {3},
pages = {881--905},
title = {{Photometric characterization of a well-defined sample of isolated galaxies in the context of the AMIGA project}},
volume = {390},
year = {2008}
}

@article{Durbala2009,
author = {Durbala, A. and Buta, R. and Sulentic, J. W. and Verdes-Montenegro, L.},
doi = {10.1111/j.1365-2966.2009.15051.x},
journal = {\mnras},
month = {aug},
number = {4},
pages = {1756--1775},
title = {{Fourier photometric analysis of isolated galaxies in the context of the AMIGA project}},
volume = {397},
year = {2009}
}

@article{Elmegreen1985,
author = {Elmegreen, B. G. and Elmegreen, D. M.},
doi = {10.1086/162810},
journal = {\apj},
month = {jan},
pages = {438},
title = {{Properties of barred spiral galaxies}},
volume = {288},
year = {1985}
}

@ARTICLE{Emsellem2006,
author = {{Emsellem}, Eric and {Fathi}, Kambiz and {Wozniak}, Herv{\'e} and {Ferruit}, Pierre and {Mundell}, Carole G. and {Schinnerer}, Eva},
title = "{Gas and stellar dynamics in NGC 1068: probing the galactic gravitational potential}",
journal = {\mnras},
year = 2006,
month = jan,
volume = {365},
number = {2},
pages = {367-384},
doi = {10.1111/j.1365-2966.2005.09716.x},
}

@article{Erwin2005,
author = {Erwin, P.},
doi = {10.1111/j.1365-2966.2005.09560.x},
journal = {\mnras},
month = {nov},
number = {1},
pages = {283--302},
title = {{How large are the bars in barred galaxies?}},
volume = {364},
year = {2005}
}

@article{Erwin2019,
author = {Erwin, Peter},
doi = {10.1093/mnras/stz2363},
journal = {\mnras},
month = {nov},
number = {3},
pages = {3553--3564},
title = {{What determines the sizes of bars in spiral galaxies?}},
volume = {489},
year = {2019}
}

@article{Eyre2011,
author = {Eyre, Andy and Binney, James},
doi = {10.1111/j.1365-2966.2011.18270.x},
journal = {Monthly Notices of the Royal Astronomical Society},
month = {may},
number = {3},
pages = {1852--1874},
title = {{The mechanics of tidal streams}},
volume = {413},
year = {2011}
}

@article{Fluke2020,
author = {Fluke, Christopher J. and Jacobs, Colin},
doi = {10.1002/widm.1349},
journal = {WIREs Data Mining and Knowledge Discovery},
month = {mar},
number = {2},
title = {{Surveying the reach and maturity of machine learning and artificial intelligence in astronomy}},
volume = {10},
year = {2020}
}

@ARTICLE{Fortson2018,
author = {{Fortson}, Lucy and {Wright}, Darryl and {Lintott}, Chris and {Trouille}, Laura},
title = "{Optimizing the Human-Machine Partnership with Zooniverse}",
journal = {arXiv e-prints},
year = 2018,
month = sep,
eid = {arXiv:1809.09738},
pages = {arXiv:1809.09738},
doi = {10.48550/arXiv.1809.09738}
}

@ARTICLE{Foster2025,
author = {{Foster}, Caroline and {Donoghoe}, Mark and {Battisti}, Andrew and {D'Eugenio}, Francesco and {Harborne}, Katherine and {Venville}, Thomas and {Lagos}, Claudia and {Mendel}, Jon Trevor and {Bagge}, Ryan and {Barsanti}, Stefania and {Bellstedt}, Sabine and {Boecker}, Alina and {Chen}, Qianhui and {Derkenne}, Caro and {Ferr{\'e}-Matteu}, Anna and {Gjergo}, Eda and {Gupta}, Anshu and {Muller}, Eric and {Santucci}, Giulia and {Park}, Hye-Jin and {Remus}, Rhea-Silvia and {Thater}, Sabine and {van de Sande}, Jesse and {Vaughan}, Sam P. and {Brough}, Sarah and {Croom}, Scott and {Valenzuela}, Lucas and {Wisnioski}, Emily and {The Magpi Team}},
title = "{The MAGPI Survey: The kinematic morphology{\textendash}density relation (or lack thereof) and the Hubble sequence at z {\ensuremath{\sim}} 0.3}",
journal = {\pasa},
year = 2025,
month = mar,
volume = {42},
eid = {e058},
pages = {e058},
doi = {10.1017/pasa.2025.19}
}

@ARTICLE{Friedli1993,
author = {{Friedli}, D. and {Benz}, W.},
title = "{Secular evolution of isolated barred galaxies. I. Gravitational coupling between stellar bars and interstellar medium.}",
journal = {\aap},
year = 1993,
month = feb,
volume = {268},
pages = {65-85}
}

@article{Fraser-McKelvie2020,
author = {Fraser-McKelvie, Amelia and Merrifield, Michael and Arag{\'{o}}n-Salamanca, Alfonso and Peterken, Thomas and Kraljic, Katarina and Masters, Karen and Stark, David and Fragkoudi, Francesca and Smethurst, Rebecca and {Fraser Boardman}, Nicholas and Drory, Niv and Lane, Richard R},
doi = {10.1093/mnras/staa2866},
journal = {\mnras},
month = {oct},
number = {1},
pages = {1116--1125},
title = {{SDSS-IV MaNGA: The link between bars and the early cessation of star formation in spiral galaxies}},
volume = {499},
year = {2020}
}

@ARTICLE{Fraser-McKelvie2025,
author = {{Fraser-McKelvie}, A. and {Gadotti}, D.~A. and {Fragkoudi}, F. and {de S{\'a}-Freitas}, C. and {Martig}, M. and {Bureau}, M. and {Davis}, T. and {Elliott}, R. and {Emsellem}, E. and {Fisher}, D. and {Hayden}, M.~R. and {van de Sande}, J. and {Watts}, A.~B.},
title = "{The GECKOS Survey: revealing the formation history of a barred galaxy via structural decomposition and resolved spectroscopy}",
journal = {arXiv e-prints},
year = 2025,
month = sep,
eid = {arXiv:2509.15976},
pages = {arXiv:2509.15976},
doi = {10.48550/arXiv.2509.15976},
archivePrefix = {arXiv},
eprint = {2509.15976},
primaryClass = {astro-ph.GA}
}

@ARTICLE{Frosst2025,
author = {{Frosst}, Matthew and {Obreschkow}, Danail and {Ludlow}, Aaron and {Bottrell}, Connor and {Genel}, Shy},
title = "{The complex relationship between black hole feedback, star formation, and stellar bars in TNG50}",
journal = {\mnras},
year = 2025,
month = mar,
volume = {537},
number = {4},
pages = {3543-3552},
doi = {10.1093/mnras/staf255}
}

@article{Gadotti2006,
author = {Gadotti, D. A. and de Souza, R. E.},
doi = {10.1086/500175},
journal = {\apjs},
month = {apr},
number = {2},
pages = {270--281},
title = {{On the Lengths, Colors, and Ages of 18 Face‐on Bars}},
volume = {163},
year = {2006}
}

@article{Gadotti2008,
author = {Gadotti, Dimitri Alexei},
doi = {10.1111/j.1365-2966.2007.12723.x},
journal = {\mnras},
month = {feb},
number = {1},
pages = {420--439},
title = {{Image decomposition of barred galaxies and AGN hosts}},
volume = {384},
year = {2008}
}

@article{Garcia-Gomez2017,
author = {Garcia-G{\'{o}}mez, C. and Athanassoula, E. and Barber{\`{a}}, C. and Bosma, A.},
doi = {10.1051/0004-6361/201628830},
journal = {\aap},
month = {may},
pages = {A132},
title = {{Measuring bar strength using Fourier analysis of galaxy images}},
volume = {601},
year = {2017}
}

@article{Garland2023,
author = {Garland, Izzy L and Fahey, Matthew J and Simmons, Brooke D and Smethurst, Rebecca J and Lintott, Chris J and Shanahan, Jesse and Silcock, Maddie S and Smith, Joshua and Keel, William C and Coil, Alison and G{\'{e}}ron, Tobias and Kruk, Sandor and Masters, Karen L and O'Ryan, David and Thorne, Matthew R and Wiersema, Klaas},
doi = {10.1093/mnras/stad966},
journal = {\mnras},
month = {apr},
number = {1},
pages = {211--225},
title = {{The most luminous, merger-free AGNs show only marginal correlation with bar presence}},
volume = {522},
year = {2023}
}

@article{Geron2021,
author = {G{\'{e}}ron, Tobias and Smethurst, R J and Lintott, Chris and Kruk, Sandor and Masters, Karen L and Simmons, Brooke and Stark, David V},
doi = {10.1093/mnras/stab2064},
journal = {\mnras},
month = {sep},
number = {3},
pages = {4389--4408},
title = {{Galaxy zoo: stronger bars facilitate quenching in star-forming galaxies}},
volume = {507},
year = {2021}
}

@ARTICLE{Geron2024,
author = {{G{\'e}ron}, Tobias and {Smethurst}, R.~J. and {Lintott}, Chris and {Masters}, Karen L. and {Garland}, I.~L. and {Mengistu}, Petra and {O'Ryan}, David and {Simmons}, B.~D.},
title = "{The Effects of Bar Strength and Kinematics on Galaxy Evolution: Slow Strong Bars Affect Their Hosts the Most}",
journal = {\apj},
year = 2024,
month = oct,
volume = {973},
number = {2},
eid = {129},
pages = {129},
doi = {10.3847/1538-4357/ad66b7}
}

@ARTICLE{Guo2023,
author = {{Guo}, Yuchen and {Jogee}, Shardha and {Finkelstein}, Steven L. and {Chen}, Zilei and {Wise}, Eden and {Bagley}, Micaela B. and {Barro}, Guillermo and {Wuyts}, Stijn and {Kocevski}, Dale D. and {Kartaltepe}, Jeyhan S. and {McGrath}, Elizabeth J. and {Ferguson}, Henry C. and {Mobasher}, Bahram and {Giavalisco}, Mauro and {Lucas}, Ray A. and {Zavala}, Jorge A. and {Lotz}, Jennifer M. and {Grogin}, Norman A. and {Huertas-Company}, Marc and {Vega-Ferrero}, Jes{\'u}s and {Hathi}, Nimish P. and {Arrabal Haro}, Pablo and {Dickinson}, Mark and {Koekemoer}, Anton M. and {Papovich}, Casey and {Pirzkal}, Nor and {Yung}, L.~Y. Aaron and {Backhaus}, Bren E. and {Bell}, Eric F. and {Calabr{\`o}}, Antonello and {Cleri}, Nikko J. and {Coogan}, Rosemary T. and {Cooper}, M.~C. and {Costantin}, Luca and {Croton}, Darren and {Davis}, Kelcey and {Dekel}, Avishai and {Franco}, Maximilien and {Gardner}, Jonathan P. and {Holwerda}, Benne W. and {Hutchison}, Taylor A. and {Pandya}, Viraj and {P{\'e}rez-Gonz{\'a}lez}, Pablo G. and {Ravindranath}, Swara and {Rose}, Caitlin and {Trump}, Jonathan R. and {de la Vega}, Alexander and {Wang}, Weichen},
title = "{First Look at z > 1 Bars in the Rest-frame Near-infrared with JWST Early CEERS Imaging}",
journal = {\apjl},
year = 2023,
month = mar,
volume = {945},
number = {1},
eid = {L10},
pages = {L10},
doi = {10.3847/2041-8213/acacfb},
}

@article{Heidl2013,
title = {Machine learning based analysis of gender differences in visual inspection decision making},
journal = {Information Sciences},
volume = {224},
pages = {62-76},
year = {2013},
issn = {0020-0255},
doi = {https://doi.org/10.1016/j.ins.2012.09.054},
author = {Wolfgang Heidl and Stefan Thumfart and Edwin Lughofer and Christian Eitzinger and Erich Peter Klement}
}

@article{Helmi1999,
author = {Helmi, A. and White, S. D. M.},
doi = {10.1046/j.1365-8711.1999.02616.x},
journal = {Monthly Notices of the Royal Astronomical Society},
month = {aug},
number = {3},
pages = {495--517},
title = {{Building up the stellar halo of the Galaxy}},
volume = {307},
year = {1999}
}

@article{Hoyle2011,
author = {Hoyle, Ben and Masters, Karen. L. and Nichol, Robert C. and Edmondson, Edward M. and Smith, Arfon M. and Lintott, Chris and Scranton, Ryan and Bamford, Steven and Schawinski, Kevin and Thomas, Daniel},
doi = {10.1111/j.1365-2966.2011.18979.x},
journal = {\mnras},
month = {aug},
number = {4},
pages = {3627--3640},
title = {{Galaxy Zoo: bar lengths in local disc galaxies}},
volume = {415},
year = {2011}
}

@article{Huertas-Company2023,
author = {Huertas-Company, M. and Lanusse, F.},
doi = {10.1017/pasa.2022.55},
journal = {\pasa},
month = {jan},
pages = {e001},
title = {{The Dawes Review 10: The impact of deep learning for the analysis of galaxy surveys}},
volume = {40},
year = {2023}
}

@article{Jiang2018,
author = {Jiang, Dongfei and Liu, F. S. and Zheng, Xianzhong and Yesuf, Hassen M. and Koo, David C. and Faber, S. M. and Guo, Yicheng and Koekemoer, Anton M. and Wang, Weichen and Fang, Jerome J. and Barro, Guillermo and Jia, Meng and Tong, Wei and Liu, Lu and Meng, Xianmin and Kocevski, Dale and McGrath, Elizabeth J. and Hathi, Nimish P.},
doi = {10.3847/1538-4357/aaa5ad},
journal = {\apj},
month = {feb},
number = {1},
pages = {70},
title = {{The Isophotal Structure of Star-forming Galaxies at 0.5 {\textless} z {\textless} 1.8 in CANDELS: Implications for the Evolution of Galaxy Structure}},
volume = {854},
year = {2018}
}

@article{Jogee2005,
author = {Jogee, Shardha and Scoville, Nick and Kenney, Jeffrey D. P.},
doi = {10.1086/432106},
journal = {\apj},
month = {sep},
number = {2},
pages = {837--863},
title = {{The Central Region of Barred Galaxies: Molecular Environment, Starbursts, and Secular Evolution}},
volume = {630},
year = {2005}
}

@article{Kalnajs1977,
author = {Kalnajs, A. J.},
doi = {10.1086/155086},
journal = {The Astrophysical Journal},
month = {mar},
pages = {637--644},
title = {{Dynamics of flat galaxies. IV - The integral equation for normal modes in matrix form}},
volume = {212},
year = {1977}
}

@ARTICLE{Katsioli2026,
author = {{Katsioli}, S. and {Xilouris}, E.~M. and {Galliano}, F. and {Adam}, R. and {Ade}, P. and {Ajeddig}, H. and {Amarantidis}, S. and {Andr{\'e}}, P. and {Aussel}, H. and {Baes}, M. and {Beelen}, A. and {Beno{\^\i}t}, A. and {Berta}, S. and {Bongiovanni}, A. and {Bounmy}, J. and {Bourrion}, O. and {Calvo}, M. and {Catalano}, A. and {Ch{\'e}rouvrier}, D. and {De Looze}, I. and {De Petris}, M. and {D{\'e}sert}, F.-X. and {Doyle}, S. and {Driessen}, E.~F.~C. and {Ejlali}, G. and {Ferragamo}, A. and {Gomez}, A. and {Goupy}, J. and {Hanser}, C. and {Hughes}, A. and {Jones}, A.~P. and {K{\'e}ruzor{\'e}}, F. and {Kramer}, C. and {Ladjelate}, B. and {Lagache}, G. and {Leclercq}, S. and {Lestrade}, J.-F. and {Mac{\'\i}as-P{\'e}rez}, J.~F. and {Madden}, S.~C. and {Maury}, A. and {Mayet}, F. and {Monfardini}, A. and {Moyer-Anin}, A. and {Mu{\~n}oz-Echeverr{\'\i}a}, M. and {Myserlis}, I. and {Nersesian}, A. and {Paliwal}, A. and {Pantoni}, L. and {Paradis}, D. and {Perotto}, L. and {Pisano}, G. and {Ponthieu}, N. and {Rev{\'e}ret}, V. and {Rigby}, A.~J. and {Ritacco}, A. and {Roussel}, H. and {Ruppin}, F. and {S{\'a}nchez-Portal}, M. and {Savorgnano}, S. and {Schuster}, K. and {Sievers}, A. and {Smith}, M.~W.~L. and {Tabatabaei}, F. and {Tedros}, J. and {Tucker}, C. and {Ysard}, N. and {Zylka}, R.},
title = "{Resolved ISM properties and scaling relations in the barred galaxy NGC 3627: constraints from NIKA2 observations}",
journal = {\mnras},
year = 2026,
month = jan,
doi = {10.1093/mnras/stag083}
}

@article{Katz1996,
author = {Katz, Neal and Weinberg, David H. and Hernquist, Lars},
doi = {10.1086/192305},
issn = {0067-0049},
journal = {The Astrophysical Journal Supplement Series},
month = {jul},
pages = {19},
title = {{Cosmological Simulations with TreeSPH}},
url = {http://adsabs.harvard.edu/doi/10.1086/192305},
volume = {105},
year = {1996}
}

@article{Kawata2021,
author = {Kawata, Daisuke and Baba, Junichi and Hunt, Jason A S and Sch{\"{o}}nrich, Ralph and Ciucă, Ioana and Friske, Jennifer and Seabroke, George and Cropper, Mark},
doi = {10.1093/mnras/stab2582},
journal = {Monthly Notices of the Royal Astronomical Society},
month = {sep},
number = {1},
pages = {728--736},
title = {{Galactic bar resonances inferred from kinematically hot stars in Gaia EDR3}},
volume = {508},
year = {2021}
}

@article{Keller2014,
author = {Keller, B. W. and Wadsley, J. and Benincasa, S. M. and Couchman, H. M. P.},
doi = {10.1093/mnras/stu1058},
journal = {Monthly Notices of the Royal Astronomical Society},
month = {aug},
number = {4},
pages = {3013--3025},
title = {{A superbubble feedback model for galaxy simulations}},
volume = {442},
year = {2014}
}

@article{Kim2021,
author = {Kim, Taehyun and Athanassoula, E. and Sheth, Kartik and Bosma, Albert and Park, Myeong-Gu and Lee, Yun Hee and Ann, Hong Bae},
doi = {10.3847/1538-4357/ac2300},
journal = {\apj},
month = {dec},
number = {2},
pages = {196},
title = {{Cosmic Evolution of Barred Galaxies up to z ∼ 0.84}},
volume = {922},
year = {2021}
}

@article{Kormendy1979,
author = {Kormendy, J.},
doi = {10.1086/156782},
journal = {\apj},
month = {feb},
pages = {714},
title = {{A morphological survey of bar, lens, and ring components in galaxies Secular evolution in galaxy structure}},
volume = {227},
year = {1979}
}

@article{Kormendy2004,
author = {Kormendy, John and Kennicutt, Robert C.},
doi = {10.1146/annurev.astro.42.053102.134024},
journal = {\araa},
month = {sep},
number = {1},
pages = {603--683},
title = {{Secular Evolution and the Formation of Pseudobulges in Disk Galaxies}},
volume = {42},
year = {2004}
}

@article{Kruk2018,
author = {Kruk, Sandor J. and Lintott, Chris J. and Bamford, Steven P. and Masters, Karen L. and Simmons, Brooke D. and H{\"{a}}u{\ss}ler, Boris and Cardamone, Carolin N. and Hart, Ross E. and Kelvin, Lee and Schawinski, Kevin and Smethurst, Rebecca J. and Vika, Marina},
doi = {10.1093/mnras/stx2605},
journal = {\mnras},
month = {feb},
number = {4},
pages = {4731--4753},
title = {{Galaxy Zoo: secular evolution of barred galaxies from structural decomposition of multiband images}},
volume = {473},
year = {2018}
}

@article{Kuno2000,
author = {Kuno, Nario and Nishiyama, Kohta and Nakai, Naomasa and Sorai, Kazuo and Vila-Vilar{\'{o}}, Baltasar and Handa, Toshihiro},
doi = {10.1093/pasj/52.5.775},
issn = {0004-6264},
journal = {Publications of the Astronomical Society of Japan},
month = {oct},
number = {5},
pages = {775--783},
title = {{Distribution and Kinematics of Molecular Gas in Barred Spiral Galaxies. I. NGC 3504}},
volume = {52},
year = {2000}
}

@article{Lahav1995,
author = {{Lahav}, O. and {Naim}, A. and {Buta}, R.~J. and {Corwin}, H.~G. and {de Vaucouleurs}, G. and {Dressler}, A. and {Huchra}, J.~P. and {van den Bergh}, S. and {Raychaudhury}, S. and {Sodre}, Jr., L. and {Storrie-Lombardi}, M.~C.},
title = "{Galaxies, Human Eyes, and Artificial Neural Networks}",
journal = {Science},
year = 1995,
month = feb,
volume = {267},
number = {5199},
pages = {859-862},
doi = {10.1126/science.267.5199.859}
}

@article{Laine2002,
author = {Laine, Seppo and Shlosman, Isaac and Knapen, Johan H. and Peletier, Reynier F.},
doi = {10.1086/323964},
journal = {\apj},
month = {mar},
number = {1},
pages = {97--117},
title = {{Nested and Single Bars in Seyfert and Non‐Seyfert Galaxies}},
volume = {567},
year = {2002}
}

@ARTICLE{LeConte2024,
author = {{Le Conte}, Zoe A. and {Gadotti}, Dimitri A. and {Ferreira}, Leonardo and {Conselice}, Christopher J. and {de S{\'a}-Freitas}, Camila and {Kim}, Taehyun and {Neumann}, Justus and {Fragkoudi}, Francesca and {Athanassoula}, E. and {Adams}, Nathan J.},
title = "{A JWST investigation into the bar fraction at redshifts 1 {\ensuremath{\leq}} z {\ensuremath{\leq}} 3}",
journal = {\mnras},
year = 2024,
month = may,
volume = {530},
number = {2},
pages = {1984-2000},
doi = {10.1093/mnras/stae921},
}

@ARTICLE{LeConte2025,
author = {{Le Conte}, Zoe A. and {Gadotti}, Dimitri A. and {Ferreira}, Leonardo and {Conselice}, Christopher J. and {de S{\'a}-Freitas}, Camila and {Kim}, Taehyun and {Neumann}, Justus and {Fragkoudi}, Francesca and {Athanassoula}, E. and {Adams}, Nathan J.},
title = "{The evolution of the bar fraction and bar lengths in the last 12 billion years}",
journal = {\mnras},
year = 2025,
month = nov,
doi = {10.1093/mnras/staf2010},
}

@ARTICLE{Liang2024,
author = {{Liang}, Xinyue and {Yu}, Si-Yue and {Fang}, Taotao and {Ho}, Luis C.},
title = "{The robustness in identifying and quantifying high-redshift bars using JWST observations}",
journal = {\aap},
year = 2024,
month = aug,
volume = {688},
eid = {A158},
pages = {A158},
doi = {10.1051/0004-6361/202348539},
}

@INPROCEEDINGS{Lindblad1994,
author = {{Lindblad}, Per Olof and {Lindblad}, Per A.~B.},
title = "{Kinematics of Interstellar Matter in a Barred Potential in the Epicyclic Approximation}",
booktitle = {Physics of the Gaseous and Stellar Disks of the Galaxy},
year = 1994,
editor = {{King}, Ivan R.},
series = {Astronomical Society of the Pacific Conference Series},
volume = {66},
month = jan,
pages = {29}
}

@ARTICLE{Lucey2023,
author = {{Lucey}, Madeline and {Pearson}, Sarah and {Hunt}, Jason A.~S. and {Hawkins}, Keith and {Ness}, Melissa and {Petersen}, Michael S. and {Price-Whelan}, Adrian M. and {Weinberg}, Martin D.},
title = "{Dynamically constraining the length of the Milky way bar}",
journal = {\mnras},
year = 2023,
month = apr,
volume = {520},
number = {3},
pages = {4779-4792},
doi = {10.1093/mnras/stad406}
}

@article{Lynden-Bell1972,
author = {Lynden-Bell, D. and Kalnajs, A. J.},
doi = {10.1093/mnras/157.1.1},
journal = {Monthly Notices of the Royal Astronomical Society},
month = {apr},
number = {1},
pages = {1--30},
title = {{On the Generating Mechanism of Spiral Structure}},
volume = {157},
year = {1972}
}

@article{Machado2016,
author = {Machado, R. E. G. and Manos, T.},
doi = {10.1093/mnras/stw572},
journal = {Monthly Notices of the Royal Astronomical Society},
month = {jun},
number = {4},
pages = {3578--3591},
title = {{Chaotic motion and the evolution of morphological components in a time-dependent model of a barred galaxy within a dark matter halo}},
volume = {458},
year = {2016}
}

@ARTICLE{Maeda2025,
author = {{Maeda}, Fumiya and {Ohta}, Kouji and {Egusa}, Fumi and {Fujimoto}, Yusuke and {Kobayashi}, Masato I.~N. and {Inoue}, Shin and {Habe}, Asao},
title = "{Galactic Structure Dependence of Cloud{\textendash}Cloud-collision-driven Star Formation in the Barred Galaxy NGC 3627}",
journal = {\apj},
year = 2025,
month = mar,
volume = {981},
number = {2},
eid = {156},
pages = {156},
doi = {10.3847/1538-4357/adb41e},
}

@article{Malhan2022,
author = {Malhan, Khyati and Ibata, Rodrigo A. and Sharma, Sanjib and Famaey, Benoit and Bellazzini, Michele and Carlberg, Raymond G. and D'Souza, Richard and Yuan, Zhen and Martin, Nicolas F. and Thomas, Guillaume F.},
doi = {10.3847/1538-4357/ac4d2a},
journal = {The Astrophysical Journal},
month = {feb},
number = {2},
pages = {107},
title = {{The Global Dynamical Atlas of the Milky Way Mergers: Constraints from Gaia EDR3–based Orbits of Globular Clusters, Stellar Streams, and Satellite Galaxies}},
volume = {926},
year = {2022}
}

@article{Marinova2007,
author = {Marinova, Irina and Jogee, Shardha},
doi = {10.1086/512355},
journal = {\apj},
month = {apr},
number = {2},
pages = {1176--1197},
title = {{Characterizing Bars at z ∼ 0 in the Optical and NIR: Implications for the Evolution of Barred Disks with Redshift}},
volume = {659},
year = {2007}
}

@article{Martinet1981,
author = {Martinet, L. and Magnenat, P. and Verhulst, F.},
doi = {10.1007/BF01301811},
issn = {0008-8714},
journal = {Celestial Mechanics},
month = {sep},
number = {1},
pages = {93--99},
title = {{On the number of isolating integrals in resonant systems with 3 degrees of freedom}},
volume = {25},
year = {1981}
}

@article{Masters2011,
author = {Masters, Karen L. and Nichol, Robert C. and Hoyle, Ben and Lintott, Chris and Bamford, Steven P. and Edmondson, Edward M. and Fortson, Lucy and Keel, William C. and Schawinski, Kevin and Smith, Arfon M. and Thomas, Daniel},
doi = {10.1111/j.1365-2966.2010.17834.x},
journal = {\mnras},
month = {mar},
number = {3},
pages = {2026--2034},
title = {{Galaxy Zoo: bars in disc galaxies}},
volume = {411},
year = {2011}
}

@article{Masters2012,
author = {Masters, Karen L. and Nichol, Robert C. and Haynes, Martha P. and Keel, William C. and Lintott, Chris and Simmons, Brooke and Skibba, Ramin and Bamford, Steven and Giovanelli, Riccardo and Schawinski, Kevin},
doi = {10.1111/j.1365-2966.2012.21377.x},
journal = {\mnras},
month = {aug},
number = {3},
pages = {2180--2192},
title = {{Galaxy Zoo and ALFALFA: atomic gas and the regulation of star formation in barred disc galaxies}},
volume = {424},
year = {2012}
}

@article{Masters2021,
author = {Masters, Karen L and Krawczyk, Coleman and Shamsi, Shoaib and Todd, Alexander and Finnegan, Daniel and Bershady, Matthew and Bundy, Kevin and Cherinka, Brian and Fraser-McKelvie, Amelia and Krishnarao, Dhanesh and Kruk, Sandor and Lane, Richard R and Law, David and Lintott, Chris and Merrifield, Michael and Simmons, Brooke and Weijmans, Anne-Marie and Yan, Renbin},
doi = {10.1093/mnras/stab2282},
journal = {\mnras},
month = {sep},
number = {3},
pages = {3923--3935},
title = {{Galaxy Zoo: 3D – crowdsourced bar, spiral, and foreground star masks for MaNGA target galaxies}},
volume = {507},
year = {2021}
}

@misc{matlab,
author = {MathWorks},
title = {{MATLAB: The Language of Technical Computing}},
url = {https://www.mathworks.com/products/matlab.html?s_tid=hp_ff_p_matlab},
year = {2024}
}

@article{McMillan2011,
author = {McMillan, Paul J.},
doi = {10.1111/j.1365-2966.2011.19520.x},
journal = {Monthly Notices of the Royal Astronomical Society},
month = {dec},
number = {3},
pages = {1565--1574},
title = {{The solar neighbourhood in angle coordinates: the Hyades moving group}},
volume = {418},
year = {2011}
}

@article{Melvin2014,
author = {Melvin, Thomas and Masters, Karen and Lintott, Chris and Nichol, Robert C. and Simmons, Brooke and Bamford, Steven P. and Casteels, Kevin R. V. and Cheung, Edmond and Edmondson, Edward M. and Fortson, Lucy and Schawinski, Kevin and Skibba, Ramin A. and Smith, Arfon M. and Willett, Kyle W.},
doi = {10.1093/mnras/stt2397},
journal = {\mnras},
month = {mar},
number = {4},
pages = {2882--2897},
title = {{Galaxy Zoo: an independent look at the evolution of the bar fraction over the last eight billion years from HST-COSMOS}},
volume = {438},
year = {2014}
}

@article{Menendez-Delmestre2007,
author = {Menendez‐Delmestre, Karin and Sheth, Kartik and Schinnerer, Eva and Jarrett, Thomas H. and Scoville, Nick Z.},
doi = {10.1086/511025},
journal = {\apj},
month = {mar},
number = {2},
pages = {790--804},
title = {{A Near‐Infrared Study of 2MASS Bars in Local Galaxies: An Anchor for High‐Redshift Studies}},
volume = {657},
year = {2007}
}

@article{Mikkola2020,
author = {Mikkola, Daniel and McMillan, Paul J and Hobbs, David},
doi = {10.1093/mnras/staa1223},
journal = {Monthly Notices of the Royal Astronomical Society},
month = {jul},
number = {3},
pages = {3295--3306},
title = {{Radial migration and vertical action in N-body simulations}},
volume = {495},
year = {2020}
}

@ARTICLE{MiyamotoNagai1975,
author = {{Miyamoto}, M. and {Nagai}, R.},
title = "{Three-Dimensional Models for the Distribution of Mass in Galaxies}",
journal = {\pasj},
year = 1975,
month = dec,
volume = {27},
number = {4},
pages = {533-543},
doi = {10.1093/pasj/27.4.533}
}

@article{Momose2010,
author = {Momose, Rieko and Okumura, Sachiko K. and Koda, Jin and Sawada, Tsuyoshi},
doi = {10.1088/0004-637X/721/1/383},
journal = {\apj},
number = {1},
pages = {383--394},
title = {{Star formation efficiency in the barred spiral galaxy NGC 4303}},
volume = {721},
year = {2010}
}

@article{Monari2019,
author = {Monari, G. and Famaey, B. and Siebert, A. and Wegg, C. and Gerhard, O.},
doi = {10.1051/0004-6361/201834820},
journal = {Astronomy {\&} Astrophysics},
month = {jun},
pages = {A41},
title = {{Signatures of the resonances of a large Galactic bar in local velocity space}},
volume = {626},
year = {2019}
}

@article{Nair2010,
author = {Nair, Preethi B. and Abraham, Roberto G.},
doi = {10.1088/2041-8205/714/2/L260},
issn = {2041-8205},
journal = {\apj},
month = {may},
number = {2},
pages = {L260--L264},
title = {{ON THE FRACTION OF BARRED SPIRAL GALAXIES}},
volume = {714},
year = {2010}
}

@article{Nakada1991,
author = {Nakada, Y. and Degucji, S. and Hashimoto, O. and Izumiura, H. and Onaka, T. and Sekiguchi, K. and Yamamura, I.},
doi = {10.1038/353140a0},
journal = {Nature},
month = {sep},
number = {6340},
pages = {140--141},
title = {{Is the bulge of our Galaxy triaxial?}},
volume = {353},
year = {1991}
}

@ARTICLE{NFW1997,
author = {{Navarro}, Julio F. and {Frenk}, Carlos S. and {White}, Simon D.~M.},
title = "{A Universal Density Profile from Hierarchical Clustering}",
journal = {\apj},
year = 1997,
month = dec,
volume = {490},
number = {2},
pages = {493-508},
doi = {10.1086/304888}
}

@ARTICLE{Nilipour2024,
author = {{Nilipour}, Andy and {Ott}, Juergen and {Meier}, David S. and {Svoboda}, Brian and {Sormani}, Mattia C. and {Ginsburg}, Adam and {Gramze}, Savannah R. and {Butterfield}, Natalie O. and {Klessen}, Ralf S.},
title = "{Turbulent Pressure Heats Gas and Suppresses Star Formation in Galactic Bar Molecular Clouds}",
journal = {\apj},
year = 2024,
month = dec,
volume = {977},
number = {1},
eid = {37},
pages = {37},
doi = {10.3847/1538-4357/ad8631}
}

@article{Ohta1990,
author = {Ohta, Kouji and Hamabe, Masaru and Wakamatsu, Ken-Ichi},
doi = {10.1086/168892},
issn = {0004-637X},
journal = {\apj},
month = {jul},
pages = {71},
title = {{Surface photometry of barred spiral galaxies}},
volume = {357},
year = {1990}
}

@ARTICLE{ONeill2003,
author = {{O'Neill}, J.~K. and {Dubinski}, John},
title = "{Detailed comparison of the structures and kinematics of simulated and observed barred galaxies}",
journal = {\mnras},
year = 2003,
month = nov,
volume = {346},
number = {1},
pages = {251-264},
doi = {10.1046/j.1365-2966.2003.07085.x}
}

@article{Paczynski1994,
author = {Paczynski, B. and Stanek, K. Z. and Udalski, A. and Szymanski, M. and Kaluzny, J. and Kubiak, M. and Mateo, M. and Krzeminski, W.},
doi = {10.1086/187607},
journal = {\apj},
month = {nov},
pages = {L113},
title = {{Are the OGLE microlenses in the galactic bar?}},
volume = {435},
year = {1994}
}

@ARTICLE{Petersen2024,
author = {{Petersen}, Michael S. and {Weinberg}, Martin D. and {Katz}, Neal},
title = "{Measuring the dynamical length of galactic bars}",
journal = {\mnras},
year = 2024,
month = jun,
volume = {531},
number = {1},
pages = {751-763},
doi = {10.1093/mnras/stae736}
}

@article{Pettitt2018,
author = {Pettitt, Alex R. and Wadsley, J. W.},
doi = {10.1093/mnras/stx3129},
journal = {\mnras},
number = {4},
pages = {5645--5671},
title = {{Bars and spirals in tidal interactions with an ensemble of galaxy mass models}},
volume = {474},
year = {2018}
}

@ARTICLE{Pettitt2020,
author = {{Pettitt}, Alex R. and {Ragan}, Sarah E. and {Smith}, Martin C.},
title = "{Young stars as tracers of a barred-spiral Milky Way}",
journal = {\mnras},
year = 2020,
month = jan,
volume = {491},
number = {2},
pages = {2162-2179},
doi = {10.1093/mnras/stz3155}
}

@misc{Pontzen2013,
author = {{Pontzen}, Andrew and {Ro{\v{s}}kar}, Rok and {Stinson}, Greg and {Woods}, Rory},
title = "{pynbody: N-Body/SPH analysis for python}",
howpublished = {Astrophysics Source Code Library, record ascl:1305.002},
year = 2013,
month = may,
eid = {ascl:1305.002},
}

@article{Qian2022,
author = {Qian, Y. and Berenbaum, S.A. and Gilmore, R.O.},
doi = {10.1038/s41598-022-22269-y},
journal = {Sci Rep},
pages = {17623},
title = {{Vision contributes to sex differences in spatial cognition and activity interests}},
volume = {12},
year = {2022}
}

@article{Querejeta2021,
author = {Querejeta, M. and Schinnerer, E. and Meidt, S. and Sun, J. and Leroy, A. K. and Emsellem, E. and Klessen, R. S. and Mu{\~{n}}oz-Mateos, J. C. and Salo, H. and Laurikainen, E. and Be{\v{s}}li{\'{c}}, I. and Blanc, G. A. and Chevance, M. and Dale, D. A. and Eibensteiner, C. and Faesi, C. and Garc{\'{i}}a-Rodr{\'{i}}guez, A. and Glover, S. C. O. and Grasha, K. and Henshaw, J. and Herrera, C. and Hughes, A. and Kreckel, K. and Kruijssen, J. M. D. and Liu, D. and Murphy, E. J. and Pan, H.-A. and Pety, J. and Razza, A. and Rosolowsky, E. and Saito, T. and Schruba, A. and Usero, A. and Watkins, E. J. and Williams, T. G.},
doi = {10.1051/0004-6361/202140695},
journal = {\aap},
month = {dec},
pages = {A133},
title = {{Stellar structures, molecular gas, and star formation across the PHANGS sample of nearby galaxies}},
volume = {656},
year = {2021}
}

@article{Raha1991,
author = {Raha, N. and Sellwood, J. A. and James, R. A. and Kahn, F. D.},
doi = {10.1038/352411a0},
journal = {Nature},
month = {aug},
number = {6334},
pages = {411--412},
title = {{A dynamical instability of bars in disk galaxies}},
volume = {352},
year = {1991}
}

@article{Rauch1996,
author = {Rauch, Kevin P. and Tremaine, Scott},
doi = {10.1016/S1384-1076(96)00012-7},
journal = {New Astronomy},
month = {oct},
number = {2},
pages = {149--170},
title = {{Resonant relaxation in stellar systems}},
volume = {1},
year = {1996}
}

@article{Reese2007,
author = {Reese, A. S. and Williams, T. B. and Sellwood, J. A. and Barnes, Eric I. and Powell, Brian A.},
doi = {10.1086/516826},
journal = {\aj},
month = {jun},
number = {6},
pages = {2846--2858},
title = {{Photometric Decomposition of Barred Galaxies}},
volume = {133},
year = {2007}
}

@ARTICLE{Regan1999,
author = {{Regan}, Michael W. and {Sheth}, Kartik and {Vogel}, Stuart N.},
title = "{Molecular Gas Kinematics in Barred Spiral Galaxies}",
journal = {\apj},
year = 1999,
month = nov,
volume = {526},
number = {1},
pages = {97-113},
doi = {10.1086/307960},
}

@ARTICLE{Romeo2023,
author = {{Romeo}, Alessandro B. and {Agertz}, Oscar and {Renaud}, Florent},
title = "{The specific angular momentum of disc galaxies and its connection with galaxy morphology, bar structure, and disc gravitational instability}",
journal = {\mnras},
year = 2023,
month = jan,
volume = {518},
number = {1},
pages = {1002-1021},
doi = {10.1093/mnras/stac3074},
}

@article{Roskar2012,
author = {Ro{\v{s}}kar, Rok and Debattista, Victor P. and Quinn, Thomas R. and Wadsley, James},
doi = {10.1111/j.1365-2966.2012.21860.x},
journal = {Monthly Notices of the Royal Astronomical Society},
number = {3},
pages = {2089--2106},
title = {{Radial migration in disc galaxies-I. Transient spiral structure and dynamics}},
volume = {426},
year = {2012}
}

@ARTICLE{Rutherford2025,
author = {{Rutherford}, T.~H. and {Fraser-McKelvie}, A. and {Emsellem}, E. and {van de Sande}, J. and {Croom}, S.~M. and {Poci}, A. and {Martig}, M. and {Gadotti}, D.~A. and {Pinna}, F. and {Valenzuela}, L.~M. and {van de Ven}, G. and {Bland-Hawthorn}, J. and {Das}, P. and {Davis}, T.~A. and {Elliott}, R. and {Fisher}, D.~B. and {Hayden}, M.~R. and {Mailvaganam}, A. and {Sharma}, S. and {Zafar}, T.},
title = "{The GECKOS survey: Jeans anisotropic models of edge-on discs uncover the impact of dust and kinematic structures}",
journal = {\aap},
year = 2025,
month = nov,
volume = {703},
eid = {A206},
pages = {A206},
doi = {10.1051/0004-6361/202556418}
}

@article{Saha2018,
author = {Saha, Kanak and Elmegreen, Bruce},
doi = {10.3847/1538-4357/aabacd},
journal = {\apj},
number = {1},
pages = {24},
publisher = {IOP Publishing},
title = {{Why Are Some Galaxies Not Barred?}},
volume = {858},
year = {2018}
}

@article{Sanders2013,
author = {Sanders, Jason L. and Binney, James},
doi = {10.1093/mnras/stt816},
journal = {Monthly Notices of the Royal Astronomical Society},
month = {aug},
number = {3},
pages = {1826--1836},
title = {{Stream–orbit misalignment – II. A new algorithm to constrain the Galactic potential}},
volume = {433},
year = {2013}
}

@article{Sanders2015,
author = {Sanders, Jason L. and Binney, James},
doi = {10.1093/mnras/stu2598},
journal = {Monthly Notices of the Royal Astronomical Society},
month = {mar},
number = {3},
pages = {2479--2496},
title = {{A fast algorithm for estimating actions in triaxial potentials}},
volume = {447},
year = {2015}
}

@article{Schinnerer2002,
author = {Schinnerer, Eva and Maciejewski, Witold and Scoville, Nick and Moustakas, Leonidas A.},
doi = {10.1086/341348},
journal = {\apj},
number = {2},
pages = {826--844},
title = {{Toward the Secondary Bar: Gas Morphology and Dynamics in NGC 4303}},
volume = {575},
year = {2002}
}

@ARTICLE{Sellwood1981,
author = {{Sellwood}, J.~A.},
title = "{Bar instability and rotation curves}",
journal = {\aap},
year = 1981,
month = jun,
volume = {99},
number = {2},
pages = {362-374},
}

@article{Sellwood2002,
author = {Sellwood, J. A. and Binney, J. J.},
doi = {10.1046/j.1365-8711.2002.05806.x},
journal = {Monthly Notices of the Royal Astronomical Society},
month = {nov},
number = {3},
pages = {785--796},
title = {{Radial mixing in galactic discs}},
volume = {336},
year = {2002}
}

@article{Sellwood2012,
author = {Sellwood, J. A.},
doi = {10.1088/0004-637X/751/1/44},
journal = {The Astrophysical Journal},
month = {may},
number = {1},
pages = {44},
title = {{SPIRAL INSTABILITIES IN N -BODY SIMULATIONS. I. EMERGENCE FROM NOISE}},
volume = {751},
year = {2012}
}

@article{Shen2010,
author = {Shen, S. and Wadsley, J. and Stinson, G.},
doi = {10.1111/j.1365-2966.2010.17047.x},
journal = {Monthly Notices of the Royal Astronomical Society},
month = {sep},
number = {3},
pages = {1581--1596},
title = {{The enrichment of the intergalactic medium with adiabatic feedback - I. Metal cooling and metal diffusion}},
volume = {407},
year = {2010}
}

@article{Sheth2002,
author = {Sheth, Kartik and Vogel, Stuart N. and Regan, Michael W. and Teuben, Peter J. and Harris, Andrew I. and Thornley, Michele D.},
doi = {10.1086/343835},
journal = {\aj},
month = {nov},
number = {5},
pages = {2581--2599},
title = {{Molecular Gas and Star Formation in Bars of Nearby Spiral Galaxies}},
volume = {124},
year = {2002}
}

@article{Sheth2008,
author = {Sheth, Kartik and Elmegreen, Debra Meloy and Elmegreen, Bruce G. and Capak, Peter and Abraham, Roberto G. and Athanassoula, E. and Ellis, Richard S. and Mobasher, Bahram and Salvato, Mara and Schinnerer, Eva and Scoville, Nicholas Z. and Spalsbury, Lori and Strubbe, Linda and Carollo, Marcella and Rich, Michael and West, Andrew A.},
doi = {10.1086/524980},
journal = {\apj},
month = {mar},
number = {2},
pages = {1141--1155},
title = {{Evolution of the Bar Fraction in COSMOS: Quantifying the Assembly of the Hubble Sequence}},
volume = {675},
year = {2008}
}

@article{Shlosman1989,
author = {Shlosman, Isaac and Frank, Juhan and Begelman, Mitchell C.},
doi = {10.1038/338045a0},
issn = {0028-0836},
journal = {Nature},
month = {mar},
number = {6210},
pages = {45--47},
title = {{Bars within bars: a mechanism for fuelling active galactic nuclei}},
volume = {338},
year = {1989}
}

@article{Solway2012,
author = {Solway, Michael and Sellwood, J. A. and Sch{\"{o}}nrich, Ralph},
doi = {10.1111/j.1365-2966.2012.20712.x},
journal = {Monthly Notices of the Royal Astronomical Society},
month = {may},
number = {2},
pages = {1363--1383},
title = {{Radial migration in galactic thick discs}},
volume = {422},
year = {2012}
}

@article{Spinoso2017,
author = {Spinoso, Daniele and Bonoli, Silvia and Dotti, Massimo and Mayer, Lucio and Madau, Piero and Bellovary, Jillian},
doi = {10.1093/mnras/stw2934},
journal = {\mnras},
month = {mar},
number = {3},
pages = {3729--3740},
title = {{Bar-driven evolution and quenching of spiral galaxies in cosmological simulations}},
volume = {465},
year = {2017}
}

@ARTICLE{Thielheim1982,
author = {{Thielheim}, K.~O. and {Wolff}, H.},
title = "{A generating mechanism of spiral structure in barred galaxies}",
journal = {\mnras},
year = 1982,
month = apr,
volume = {199},
pages = {151-169},
doi = {10.1093/mnras/199.1.151},
}

@INPROCEEDINGS{Toomre1981,
author = {{Toomre}, A.},
title = "{What amplifies the spirals}",
booktitle = {Structure and Evolution of Normal Galaxies},
year = 1981,
editor = {{Fall}, S.~M. and {Lynden-Bell}, D.},
month = jan,
pages = {111-136},
}

%%%%%%%%%%%%%%%%%%%%%%%%%%%%%%%%%%%%%%%%%%%%%%%%%%

%%%%%%%%%%%%%%%%% APPENDICES %%%%%%%%%%%%%%%%%%%%%

\appendix
\section{Kinematic Maps \& a NSD}
\label{a:nsd}
In Section \ref{s:barproperties} we note a feature in the action conservation (Figure\,\ref{f:Jtot_rbint}) which might be related to a NSD in the centre of the simulated galaxy. 
This is assumed due to the 
central feature evident in the kinematic maps of this galaxy but 
has not been subjected to extensive study as yet. The relevant kinematic maps ($v_{\rm los}$,\,$h_3$) are included in this appendix for reference.  

\begin{figure}
    \centering
	\includegraphics[width=0.69\columnwidth]{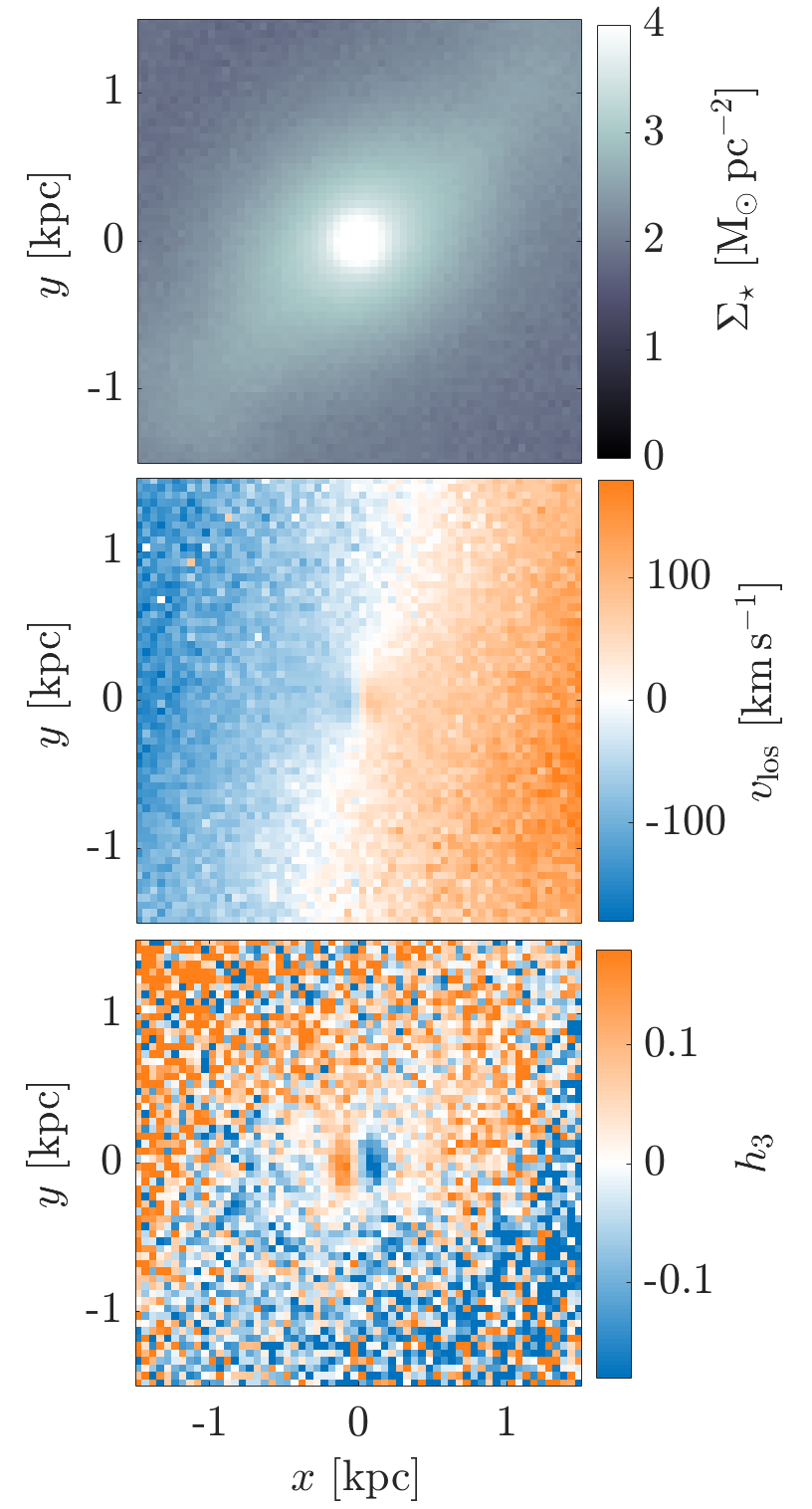}
    \caption{The face-on projected stellar density (upper), mean line-of-sight velocity ($v_{\rm los}$, middle) and skewness ($h_3$, lower) zoomed-in to the centremost ($R<1.5$\,kpc) region of the simulated galaxy at the $1$\,Gyr snapshot to highlight the potential NSD feature.
    } 
    \label{f:nsd_face}
\end{figure}

\begin{figure}
    \centering
	\includegraphics[width=0.87\columnwidth]{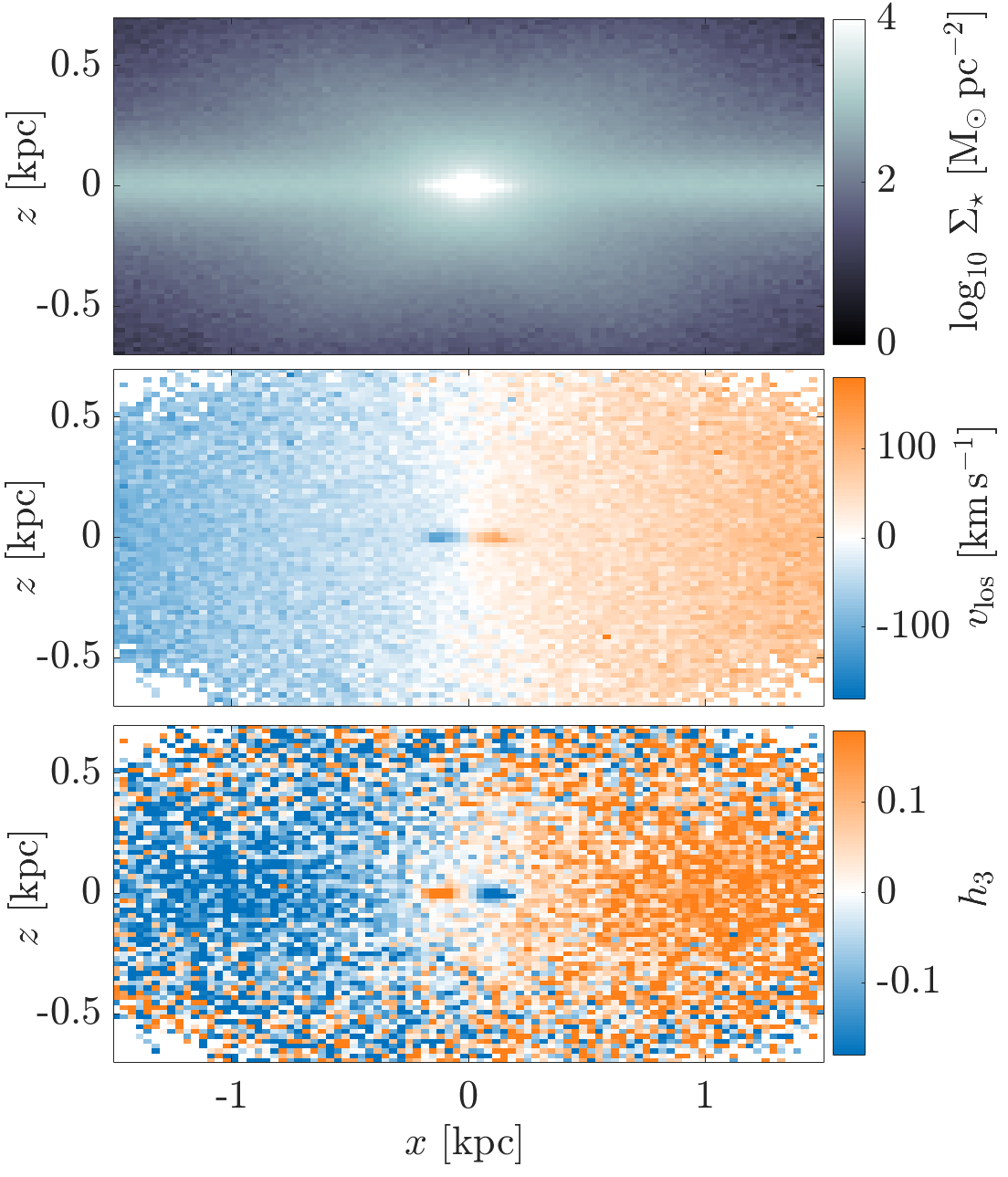}
    \caption{The edge-on projected stellar density (upper), mean line-of-sight velocity ($v_{\rm los}$, middle) and skewness ($h_3$, lower) zoomed-in to the centremost ($R<1.5$\,kpc, $z<0.6$\,kpc) region of the simulated galaxy at the $1$\,Gyr snapshot to highlight the potential NSD feature.
    }
    \label{f:nsd_edge}
\end{figure}

%%%%%%%%%%%%%%%%%%%%%%%%%%%%%%%%%%%%%%%%%%%%%%%%%%
% Don't change these lines
\bsp	% typesetting comment
\label{lastpage}
\end{document}